# 3D Imaging of Complex Skyrmion and Hopf Topologies in an Extended Sample

I. Binnie*†[1], H. Fang†[1], B. Shearer[1], A. Grafov[1], N. Jenkins[1], Y. Shao[1], C. O'Leary[7], Y. Liao[7], T. Feggeler[2,3], A. Oh[2,3], S. Yazdi[4], J. Zou[5,6], B. Wang[1], E. Cating[1], S. A. Montoya[8], D. Shapiro[3], J. Miao[7], H. C. Kapteyn[1,9], M. M. Murnane[1]

*[1] STROBE NSF Science & Technology Center and JILA, University of Colorado & NIST; Boulder, CO 80309, USA*

*[2]Department of Physics, University of California, Berkeley, Berkeley, CA 94720, United States*

*[3]Advanced Light Source, Lawrence Berkeley National Laboratory, Berkeley, CA 94720, United States*

*[4]Renewable and Sustainable Energy Institute, University of Colorado, Boulder, CO 80309, USA*

*[5]Physics Department, King Fahd University of Petroleum and Minerals, 31261, Dhahran, Saudi Arabia*

*[6]Quantum Center, KFUPM, Dhahran, Saudi Arabia*

*[7]STROBE NSF Science & Technology Center and Department of Physics & Astronomy and California NanoSystems Institute, University of California, Los Angeles, CA 90095, USA*

*[8]Center for Memory and Recording Research, University of California San Diego, La Jolla, California 92093, USA*

*[7] Kapteyn-Murnane Laboratories, Inc.; Boulder, CO 80301, USA*

**Indicates corresponding author*

*†These authors contributed equally to the work*

## Abstract

Spin textures are key for emergent magnetic phenomena such as topological protection and underpin novel spintronic device paradigms based on racetrack memory, logic gates, and neuromorphic computing. Using a coherent diffractive imaging technique called vector ptycho-tomography, in combination with algorithms that are robust to noise, we image the 3D magnetic texture of skyrmion and Hopf topologies with no prior assumptions about the sample. This directly reveals experimentally for the first time an extended 3D skyrmion lattice, including the domain wall shape, topological charge, helicity, and Hopf index. Our findings demonstrate experimentally that dipole stabilized skyrmions in Fe/Gd multilayers exhibit barrel-shaped skyrmion tubes with a twisted helicity, transitioning from Néel-type winding at the surfaces to both clockwise and counterclockwise Bloch-type winding in the bulk, that can also be described as fractional hopfions. We image a lattice of 24 skyrmions with topological charge 1, average depth-dependent domain wall width of 23 to 40 nm, depth-dependent twisted helicity from $\pm 155°$ to $\pm 30°$, and fractional Hopf index of $\pm 0.3$. Over 10 TB of data were analyzed to yield a fully-resolved 3D reconstruction over a >0.4 μm$^3$ volume, with high fidelity down to the Nyquist limit of 8 nm. This method fills a key gap in the current landscape of magnetic imaging by enabling high-resolution, element-specific 3D reconstructions of full-field extended spin textures – offering a new route for exploring the topological complexity of magnetic materials in three dimensions.

## Introduction

Nanoscale spin textures are central to the study of magnetic materials, and appear in domain walls, vortices, as well as more complex non-collinear and topologically nontrivial structures. In many magnetic systems, domain walls mediate the transition between magnetically aligned regions and play a central role in determining switching behavior, coercivity, and energy dissipation[1]. In geometrically frustrated or chiral systems, competing interactions such as exchange, Dzyaloshinskii–Moriya interactions (DMI), and dipolar coupling can give rise to complex spin arrangements, including spin spirals, magnetic bubbles, and skyrmions[2-4]. These textures are not only key to understanding emergent magnetic phenomena such as chirality, topological protection, and solitonic excitations, but they also underpin novel spintronic device paradigms based on racetrack memory, logic gates, and neuromorphic computing[5-11], and are considered a potential platform for quantum computation[12,13]. Their nanoscale structure and dynamic response to electric currents, fields, and thermal gradients make them ideal candidates for tunable, energy-efficient magnetic technologies.

Of particular interest are magnetic skyrmions, which are topologically protected, nanoscale spin configurations that behave as quasiparticles. Their stability and manipulability make them promising candidates for information storage[9] and logic operations[10] in next-generation spintronic devices. While powerful characterization methods like magnetic force microscopy (MFM) or Lorentz transmission electron microscopy (LTEM) can be used to investigate the 2D characteristics of skyrmions[14,15], their 3D spin structures remain difficult to probe directly. Instead, depth-dependent magnetic structures have been studied via indirect measurements[16-18].

Several nanoimaging techniques have recently been implemented to study the non-trivial depth-dependent structures of skyrmions. Tomographic LTEM[19] and electron holography[20,21] have provided insight into the 3D structure of confined skyrmion structures in magnetic needles and nanoplanks, where the semi-isolated sample form factor allows successful reconstruction of high-resolution vector maps. Other approaches combined electron holography with a sliced sample to extract 3D flux information[22], while neutron scattering[23] has been combined with high-resolution 2D imaging using physics-based models to extract 3D information from magnetic samples. These techniques probe 3D information through a combination of prior information about sample physics, statistical correlations, or extensive sample sectioning.

Probing static and dynamic spin textures can greatly benefit from advances in coherent light sources in the extreme ultraviolet (EUV) and soft X-ray (SXR) regions. This photon energy range spans approximately 20-2500 eV and encompasses multiple magnetic resonances for rare earth and transition metal magnetic materials, allowing for element-specific magnetic contrast via X-ray magnetic circular and linear dichroism (XMCD, XMLD). SXR wavelengths also penetrate relatively deep into materials (~1 μm) for depth-resolved imaging. Scanning transmission X-ray microscopy (STXM) has been used in combination with XMCD to perform tomography on isolated[24] or single[25] magnetic topological features with spatial localization down to 75nm and 22nm, respectively. This approach has even been extended to time-resolved studies to observe the skyrmion Hall effect[26]. SXR Fourier transform holography has also been combined with tomography to image worm domains[27] and skyrmion cocoons[28] in an extended 3D region, but the resolution limits of this technique prevent quantitative inspection of the sample topology.

Coherent diffractive imaging (CDI) has further expanded the resolution of 3D magnetic imaging, allowing topological features such as Bloch points to be experimentally observed[29]. CDI at the Advanced Light Source coherent imaging beamline 7.0.1.2 (ALS COSMIC) at Lawrence Berkely National Laboratory (LBNL) has been used to image 3D spin textures with ~10 nm spatial resolution[30] – smaller than the length scales of their topological features. That work implemented the first general 3D SXR vector ptychotomography, by imaging topological magnetic monopoles (TMMs, also known as hedgehogs or Bloch points) in a frustrated nickel metalattice sample. This sample was ideal for imaging because it consisted of Ni nanoparticles separated by ~65 nm, linked by very thin necks, with many voids. Since this sample gives rise to very strong charge (structural) as well as magnetic scattering, accurate tomographic image alignment is relatively straightforward, so that the 3D spin texture can be extracted. At the Swiss Light Source PolLux beamline, SXR laminography was employed to reconstruct the 3D spin texture of a single DMI-stabilized skyrmion with 20 nm spatial resolution, revealing depth-dependent domain size and constant helicity[31]. Since the sample was a single skyrmion disc with 800 nm diameter that was isolated within the field of view, good tomographic alignment of the images was possible.

In contrast to engineered metalattice and single-skyrmion sample geometries, the structural properties of magnetic thin film samples (such as the Fe/Gd multilayer studied here) present a unique challenge for tomographic imaging. These are particularly important samples because they support interesting topologies such as skyrmions[32-37], antiskyrmions[38], skyrmion cocoons[39,28], bloch points[25], ferroelectric domain walls[40], and hopfions[41,42], yet their charge contrast is typically low or non-existent, the magnetic features of interest often extend over large areas and can vary with depth into the sample, and the structure may not be perfectly rigid. These properties make it challenging to accurately register a series of projections in order to extract 3D spin textures.

In the case of the dipole stabilized Fe/Gd samples studied here, their 3D characteristics have been extensively studied in simulation[43,44], which predicts a hexagonal lattice of barrel-shaped skyrmion tubes with a twisted helicity, transitioning from Néel-type winding at the surfaces to both clockwise and counterclockwise Bloch-type winding in the bulk. This 3D twisted helicity can be described as a fractional hopfion, confined in its third dimension by the sample surface[44]. Direct measurement methods such as Lorentz transmission electron microscopy (LTEM) and scanning electron microscopy (SEM) have verified these simulation results in surface-sensitive and projection-based modalities[18], but a direct image of the 3D spin texture in an extended skyrmion lattice sample has yet to be demonstrated.

Here we present full-field vector-sensitive 3D nanoscale imaging of a complex magnetic texture – in this case dipole stabilized skyrmions in an Fe/Gd multilayer[43,44]. We use vector ptycho-tomography[30] with noise-robust algorithms[45] to accurately reconstruct an ~8 nm resolution 3D image from over 10 TB of diffraction data taken at the COSMIC coherent imaging beamline. This reconstruction requires no prior assumptions or knowledge about the sample. We directly extract the 3D topology of dipole-stabilized skyrmions, including depth-dependent domain wall properties, topological protection, twisted helicity, and fractional Hopf index. Our findings demonstrate experimentally that dipole stabilized skyrmions in Fe/Gd multilayers exhibit barrel-shaped skyrmion tubes with a twisted helicity, transitioning from Néel-type winding at the surfaces to both clockwise and counterclockwise Bloch-type winding in the bulk, that can also be described as a fractional hopfion. We report a lattice of skyrmions with topological charge 1, average depth-dependent domain wall width of 23 to 40 nm, depth-dependent twisted helicity of $\pm 155°$ to $\pm 30°$, and fractional Hopf index of $\pm 0.3$. Our method fills a key gap in the landscape of magnetic imaging by enabling high-resolution, element-specific 3D reconstructions of full-field extended spin textures in magnetic multilayers – offering a new route for exploring the topological complexity of magnetic materials in three dimensions. Finally, we note that this result is also an example of the large-scale data that are increasingly prevalent in modern material characterization. Methods that reliably extract physics from >TB-scale datasets are crucial to understanding complex physical systems.

### Imaging the 3D Topology of an Extended Skyrmion Lattice

Skyrmion lattices and other extended spin textures that form in thin-film multilayers present major challenges for direct imaging. Unlike structured samples such as metalattices, nanoparticles, and needles, thin-film samples have little-to-no structural contrast and instead exhibit only magnetic contrast. Extracting accurate 3D spin textures of these samples using X-ray light without prior assumptions requires a vector ptycho-tomography method[30] that is enhanced using fiducial structures (holes or nanostructures), as well as computational reconstructions that are robust in the presence of various noise sources[45]. The sample is an Fe/Gd multilayer thin-film deposited by sputtering on a SiN membrane and magnetized to support a dipole-stabilized skyrmion lattice as described in [34]. A small section is lifted out and prepared with fiducial holes via focused ion beam (FIB) and mounted onto a copper Omniprobe grid (see SI section 2), as shown in Fig. 1.

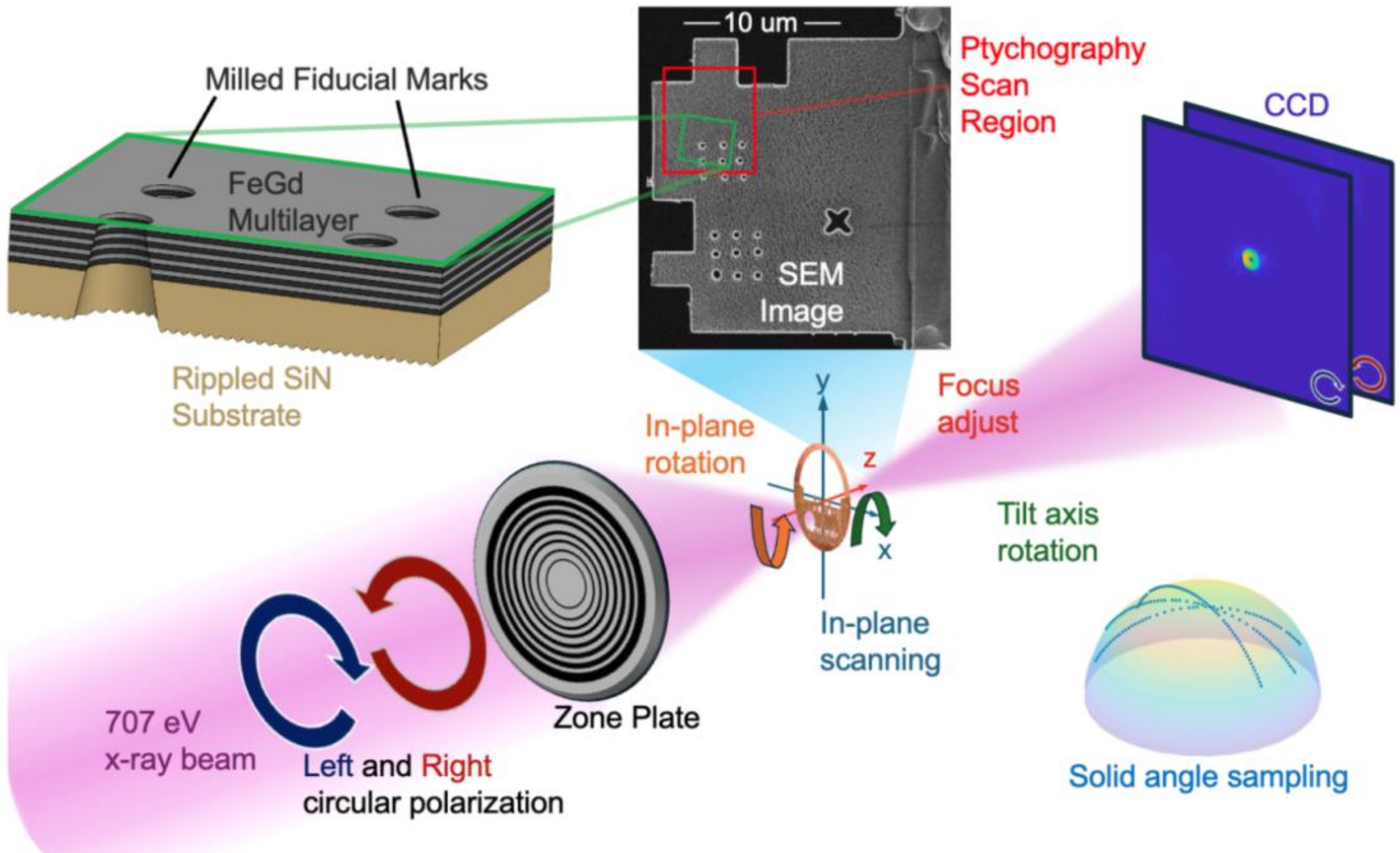

**Figure 1. 3D vector ptycho-tomography of a low-contrast skyrmion sample**. Left and right circular polarized soft X-rays beams at 707 eV are focused using a zone plate onto the sample. Scattered light from the sample is collected by a CCD, as well as background and other noise (see log scale images in Figs. 2a and b). The sample mounting allows for in-plane scanning for ptychography, $x$-axis rotation for tomography, and manual discrete $z$-axis rotation to increase the solid angle sampling (bottom right). The sample is also moved along the $z$-axis and defocused for ptychography. The inset shows an SEM image of the Fe/Gd sample and indicates the approximate ptychography scan region used

for each projection. A cartoon (top left) shows the sample together with the fiducial holes used to enable accurate alignments.

The experiment was performed at the imaging branch of the COSMIC beam line. The soft x-ray beam was tuned to the Fe $L_3$-edge (707 eV). 2D projections of the sample with both structural and magnetic contrast were taken via ptychography over a broad range of solid angles for eventual tomographic reconstruction. At each rotation, ptychographic diffraction patterns were acquired with both a left and right circularly polarized beam in order to separate the structural and magnetic signals using the XMCD effect[46] (see Fig. 1 and Methods).

For the tomographic data acquisition, three independent *x*-axis rotation series were performed from approximately -60° to 60° in 2°- 4° steps. The sample was manually rotated around the *z*-axis between each scan to increase the solid angle sampling and capture all components of the vector magnetization field, with scans performed at approximately 0°, 60°, and 210°. Reconstruction of the full 3D vector image required accurate phase retrieval and alignment of ~108,000 diffraction patterns from the ptycho-tomographic series, effectively condensing over 10 TB of data into a single nanoscale image. The reconstruction process consists of three steps: 1) ptychography to reconstruct 2D images of both left and right circular polarizations, 2) XMCD overlap to extract the structural and magnetic projections from the two polarization images, and 3) vector tomography to reconstruct the full 3D magnetic image. This process requires no physics model for or assumptions about the sample properties. Please see the Methods Section for a detailed discussion.

**Results and Discussion**

The full scalar and vector tomography reconstruction results are shown in Fig. 2, with high resolution images and movies provided in the SI, and more details given in the Methods section. We reconstruct a region approximately 2.4 μm x 2.4 μm wide and 640 nm thick with voxel size 8nm x 8nm x 8nm. The magnetic reconstruction (Fig. 2a) illustrates the magnetization distribution in the sample and constitutes the first demonstration of full-field 3D imaging of a skyrmion lattice. The structural reconstruction (Fig. 2a inset) shows well-defined fiducial holes, indicating good alignment of the projections and convergence of the tomography algorithm. A single example skyrmion feature from this reconstruction (Fig. 2b) is enlarged to demonstrate the Néel type winding at the surface layers and Bloch type winding in the central layer (Figs. 2c-e). Meanwhile, a cross section of the sample (Fig. 2f) shows dipolar magnetization field lines, in good agreement with the dipole stabilization mechanism and the skyrmion tubes obtained from micromagnetic simulations (see SI sections 5-6 for simulation details).

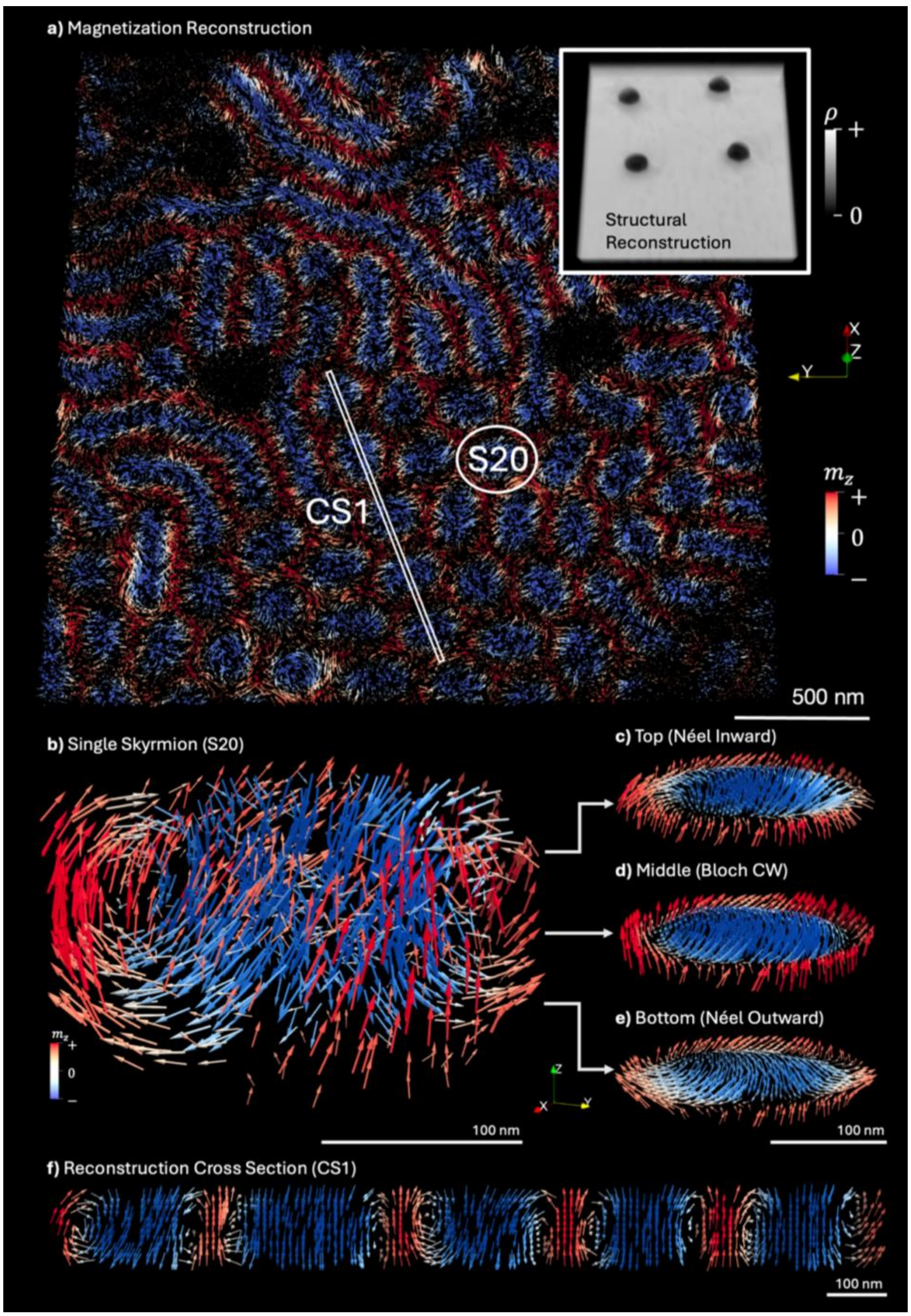

**Figure 2. 3D scalar and vector images. a)** Tomographic reconstructions of the structural signal and magnetic signal (magnetization). The structural reconstruction shows fiducial holes with clear edges, and the magnetic reconstruction shows the full-field magnetic textures featuring skyrmions and worm domains. Both reconstructions illustrate the sample in 3 dimensions over a region 2.4 μm x 2.4 μm wide. **b)** Shows a zoomed-in view of a single reconstructed

skyrmion, labeled S20 in b), and **c-e)** the top, central and bottom slices of that single skyrmion, showing Néel, Bloch, and Néel type winding, respectively. **f)** A cross section along a chain of skyrmion tubes (labeled CS1).

This reconstruction is completely independent of any simulations of the magnetic texture or physical models for its behavior. In contrast to 3D characterization methods that correlate scattering data with simulations[22,23,47] or indirectly solve for magnetic field components[20], this is a direct probe of the existing 3D magnetization texture. Fully direct methods allow for the detection of anisotropies and defects not easily simulated using micromagnetics. The only assumptions are those applied in ptychography[45], which are equivalent to physical constraints of a transmissive scattering experiment.

In addition to 3D visualization, the full-field reconstruction also accesses quantitative depth-resolved magnetic structure, calculated across many topologically protected skyrmions. Several analyses presented below reveal the depth-resolved magnetic and topological properties of an extended lattice of dipole-stabilized skyrmions.

Topological features such as skyrmions can be defined in two dimensions by their winding number

$$N = \int_A \rho \, dx \, dy, \tag{1}$$

where

$$\rho = -\frac{1}{4\pi}\left(\frac{\partial \boldsymbol{m}}{\partial x} \times \frac{\partial \boldsymbol{m}}{\partial y}\right) \cdot \boldsymbol{m} \tag{2}$$

is defined using the reconstructed magnetization vector $\boldsymbol{m}$ normalized to unit length[4]. The winding number of a 2D topologically protected feature integrates to an integer value over its area $A$ in the $xy$-plane. With this definition, an integrated $N$ value of 1 indicates a skyrmion when the core of the feature points in the $-z$ direction.

A skyrmion is theoretically defined in an infinitely extended plane with azimuthal symmetry and uniform magnetization at infinity – however, the skyrmions imaged here exist within a lattice and vary in size and shape. To define a boundary for each skyrmion, we manually select an ellipse in the xy-plane that encloses its topological charge density, while excluding charge density from other skyrmions. The skyrmion positions and sizes are relatively constant in z (see below), so the elliptical boundary is held constant throughout the reconstruction depth.

We analyze the topology of this sample in two-dimensional layers[31] and identify 24 skyrmion tubes, each with a winding number of 1 in all layers and $-z$ magnetization in the core (see Fig. 3). The persistence of the topological protection throughout the depth is consistent with micromagnetic simulations of hybrid dipole-stabilized skyrmion tubes[34,35].

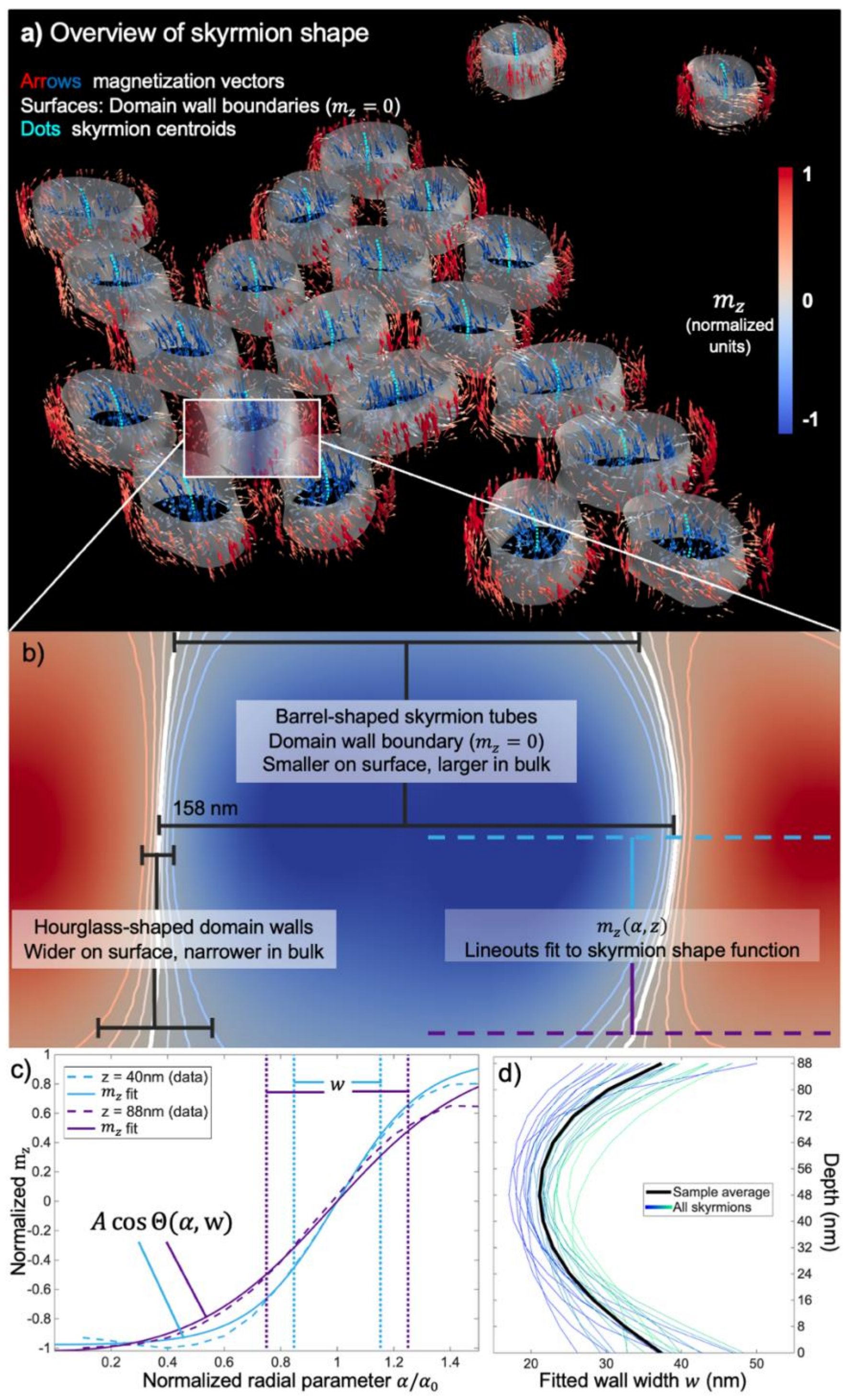

**Figure 3. Reconstructed skyrmion shapes. a)** Skyrmion features identified in this sample, with domain walls, defined as the surface where $m_z = 0$, shown in grey. The centroid of the skyrmion at each layer is shown in cyan. **b)** Example cross section of a single skyrmion, with domain wall boundaries that are broader in the bulk. The domain wall thickness, defined by the radial profile of $m_z$, is wider on the surfaces and narrower in the bulk. The dashed lines are averaged azimuthally and plotted in **c)** as a function of $\alpha/\alpha_0$, where $\alpha$ is a radial parameter defined below, and $\alpha_0 \equiv s(40\text{nm})$ (eq. [3]). The $m_z$ curve is fit to $A \cos \Theta$, with $\Theta(\alpha, w)$ given by eq. [4] (solid lines), to define the domain wall width $w$. The data and fit are normalized to the maximum of $|m_z|$. Extracted widths for each skyrmion at each depth are plotted in **d)**, with the average $w(z)$ shown in black.

The centroid position of each skyrmion can be found by performing a center-of-mass calculation on the topological charge within the boundary of each skyrmion. This method is sensitive to noise based on the region manually selected, but a less sensitive method is to calculate the center of mass of $(1 - m_z)$ within the skyrmion boundary. Both methods result in skyrmion centroids that vary in position by less than a pixel (< 8nm) through the depth of the sample, indicating that the skyrmion tubes are straight to within approximately 5% of their diameter size and parallel to the sample z-axis with a tilt angle bounded by $\tan^{-1} 8/80 = 5.7°$.

The skyrmion size can be characterized via its domain wall, which we define as the contour line of $m_z = 0$. By fitting a series of ellipses[48] to the $m_z = 0$ contour of each skyrmion at each depth, we track the depth-dependent skyrmion size $s(z)$, which is defined by the square root of product of the semi-major ($a$) and semi-minor ($b$) axes of the fitted ellipses

$$s(z) = \sqrt{a(z)b(z)}. \quad [3]$$

Although the absolute size varies, each skyrmion domain wall is larger in the bulk than on the surfaces by an average of 2nm (see SI section 4.8).

To quantify the wall thickness, we define a radial parameter for the elliptical skyrmion profile. We take the $m_z = 0$ fitted ellipse at $z = 40\ nm$ (with size $\alpha_0 \equiv s(40\text{nm})$) to set the shape of the skyrmion, then define a series of ellipses equally spaced in major and minor axis and calculate the average value of $m_z$ on each ellipse. The radial parameter $\alpha$ of each ellipse is the square root of the product of its major and minor axes. For the case of a circular skyrmion, $a = b$, and this process is equivalent to an azimuthally averaged radial profile. We fit this profile to the function $A \cos \Theta$, where $\Theta$ is given by

$$\Theta(\alpha, w) = 2 \tan^{-1}\left(\frac{\sinh \alpha_0/w}{\sinh \alpha/w}\right), \quad [4]$$

inspired by the skyrmion domain wall model in [49], with $w$ and $A$ as width and amplitude parameters, respectively. The amplitude parameter accounts for the suppression of the magnitude of $\boldsymbol{m}$ at the surfaces due to the missing wedge effect[25,50]. While this effect impacts all components of $\boldsymbol{m}$, $m_z$ is less impacted by general sampling bias (see SI section 4.9) and is therefore the best choice for quantitatively describing the domain wall behavior. All the skyrmions exhibit a domain wall width ranging from approximately 26% of the domain size (23 nm) in the center to 45% (40 nm) at the surfaces. This is a direct measurement confirmation of the simulated depth-dependent domain wall structure in dipole-stabilized skyrmions.

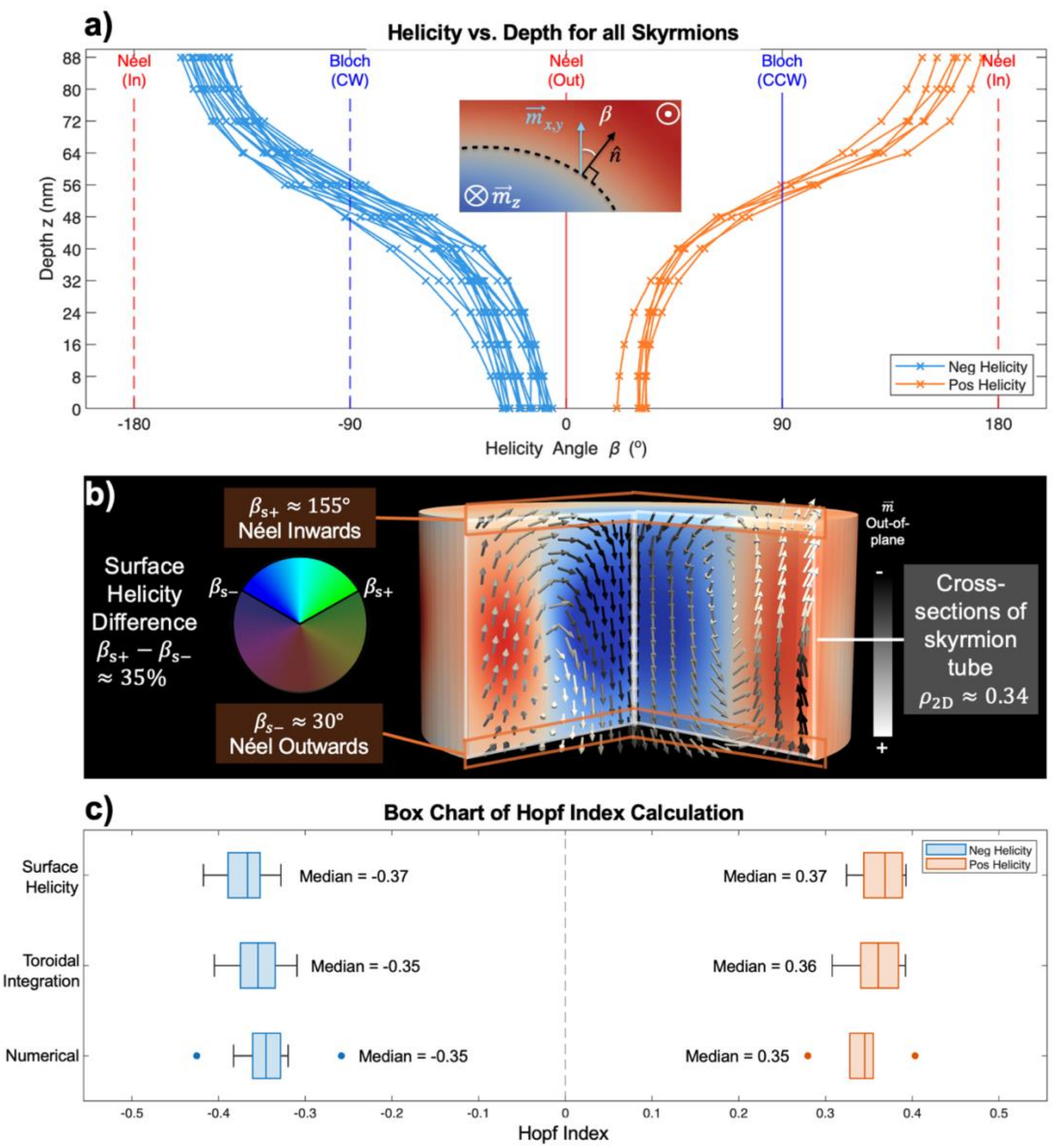

**Figure 4. Extracting the Helicity and Hopf index. a)** Helicity angle $\beta$ vs. sample depth. The inset shows $\beta$, defined as the angle between the domain wall normal and the in-plane magnetization. Each curve corresponds to a single skyrmion. $\beta$ is calculated and averaged along the domain wall of each skyrmion at each depth. Skyrmions of both helicities wind from inward Néel at the upper surface to CW or CCW Bloch at the bulk, to outward Néel at the lower surface. **b)** Illustration of two methods for calculating the Hopf index of a reconstructed skyrmion. The surface helicities $\beta_{s\pm}$ are calculated and subtracted to give the fractional winding of the skyrmion tube (left). Alternatively, an azimuthal cross-section of the skyrmion has a fractional 2D topological charge $\rho$. Integrating this cross-section toroidally results in a fractional 3D Hopf index. **c)** Hopf index calculation using the methods described above, plus a fully numerical calculation (eq. [5])[44]. The central line of each box indicates the median value across all skyrmions, the edges of the box indicate upper and lower quartile, the whiskers extend from nonoutlier minimum to nonoutlier maximum, and the dots represent outliers, which are more than 1.5*IQR away from the edges of the boxes.

While topological charge primarily describes the longitudinal winding of magnetization traveling from the skyrmion core to the outside, helicity describes the transverse winding behavior. The helicity angle $\beta$ is defined as the angle between the domain wall normal vector and the in-plane magnetization (inset of Fig. 4a). Therefore, a purely Néel skyrmion with outward (inward) winding will have $\beta = 0°\ (\pm 180°)$, and a purely Bloch skyrmion with counterclockwise (clockwise) winding will have $\beta = 90° (-90°)$[18]. Here we observe hybrid helicity skyrmions, with Néel caps on the surfaces and Bloch helicity in the central layer, in agreement with a centrosymmetric dipole-stabilized skyrmion lattice model and micromagnetic simulations[34,35]. Figure 4a shows how the helicity evolves as a function of depth, transitioning between Néel and Bloch type windings. This is the first direct observation of depth-resolved twisted helicity in a skyrmion sample.

The 3D nature of these hybrid skyrmion stacks invites a fully 3D analysis of their structure. 3D spin textures may be described by their Hopf index, which is defined as

$$H = -\int_V \boldsymbol{F} \cdot \boldsymbol{A}\, d^3\boldsymbol{r}, \qquad [5]$$

where $\boldsymbol{F}$ is the emergent magnetic field, calculated from the normalized magnetization vector field $\boldsymbol{m}$ as

$$F_i = -\frac{1}{8\pi} \epsilon_{ijk} \boldsymbol{m} \cdot \left(\partial_j \boldsymbol{m} \times \partial_k \boldsymbol{m}\right), \qquad [6]$$

and $\boldsymbol{A}$ is the vector potential of this field[51]. A magnetic Hopfion is a topologically protected spin texture with localization in all three dimensions, formed by a bi-meron tube joined end-to-end to itself in a torus. A bi-meron is topologically equivalent to a skyrmion, but has core and background magnetization that are in-plane, rather than out-of-plane[52]. The helicity completes an integer number of full circle twists ($\beta$ from 0° to 360°) around the torus, equivalent to the Hopf index.

Although the skyrmions imaged here do not exhibit localization along the z-axis (except that imposed by the sample boundaries), recent work has shown that the hybrid helicity of these dipole-stabilized features can be understood as the toroidal wrapping of merons (half-integer skyrmions), forming half-integer hopfions under Coulomb gauge[44]. It is also shown[44] that the half-integer Hopf index can be equivalently thought of as the skyrmion helicity twisting Néel-Bloch-Néel (half a circle) from one surface to the other, making the fractional Hopf index directly connected to the helicity twist along the tube. In a real physical system where the helicity at the surfaces is a hybrid of Néel and Bloch types, the Hopf index is expected to be between 0 and ±0.5 under Coulomb gauge. Quantitatively, the Hopf index can be directly linked with the difference of helicity angles at the top ($s_+$) and bottom ($s_-$) layers: $H = (\beta_{s+} - \beta_{s-})/360°$[44]. We calculate the Hopf index over each skyrmion feature in this way and compare it to the full numerical calculation carried out by numerically applying Eq. [5] to the reconstructed vector field (SI section 4.7).

A third method of calculating the Hopf index is inspired by the interpretation of a half-integer hopfion as a complete toroidal wrapping of a meron – that is, a 2D spin texture with $\rho = \pm 0.5$ – around the z-axis. We integrate the 2D topological charge density of the radial cross section of the skyrmion tube toroidally around the azimuthal angle, following the meron torus modeling (see Fig. 4b). The small variation between different calculation methods is due to the unlocalized boundary condition and the fact that each skyrmion is not fully isolated from other skyrmions under zero external field. A similar discrepancy occurs on a simulated sample with variable skyrmion shape.

The Hopf index calculations on this sample provide a verification of the predicted hybrid partial integer hopfions in this sample. They confirm, as suggested in [44], that the Hopf index of dipole-stabilized skyrmions may be calculated from a surface-sensitive measurement if the sample meets certain symmetry requirements. The 3D reconstruction can verify the complete texture of the features, as well as identify any asymmetries that may exist.

Finally, we note that the pixel size of the 2D ptychography reconstructions is $\lambda z/D$, where $z$ is the sample to detector distance and D is the detector size. This reconstruction has a pixel size of 8 nm, though the minimum pixel size with no detector cropping would be 4 nm. The diffraction patterns were cropped during preprocessing to a smaller region that did not truncate any of the scattering signal. A Fourier shell correlation analysis (see SI section 7) confirms that all three dimensions of the reconstruction are well resolved out to the Nyquist limit. We conclude that the resolution of this technique is more than sufficient to characterize the magnetic texture of the Fe/Gd multilayer.

## Conclusion

With no prior assumptions about the sample properties, we directly image the 3D magnetic textures—including the domain wall shape, topological charge, helicity, and Hopf index—of a skyrmion lattice in an extended thin film. We image 24 skyrmions, each with topological charge 1, average depth-dependent domain wall width of 23 to 40 nm, depth-dependent twisted helicity from $\pm 155°$ to $\pm 30°$, and fractional Hopf index of $\pm 0.3$. No other technique to date can image a large field of skyrmions with sensitivity to both surface and depth-resolved bulk behavior. The lateral and depth resolution is sufficient to resolve the 3D winding behavior of skyrmions and can be applied to general thin-film samples with finer magnetic features and complex topologies.

## Methods

**1) Ptychography:** Each 2D image is collected as a series of diffraction patterns from overlapping areas of the sample and reconstructed via iterative phase retrieval. For this experiment we used PaCMAN, a recently developed ptychography algorithm that can accommodate the significant high-frequency angle-dependent parasitic scattering present in the diffraction data (see Figs. 5a-b) and produce a high-fidelity diffraction-limited image[45].

Each image was then converted from transmission to optical density (OD) by taking the logarithm of the transmission of a vacuum region divided by the pixel-dependent transmission. Performing this normalization separately for each image accounts for small fluctuations in beam intensity between ptychography scans.

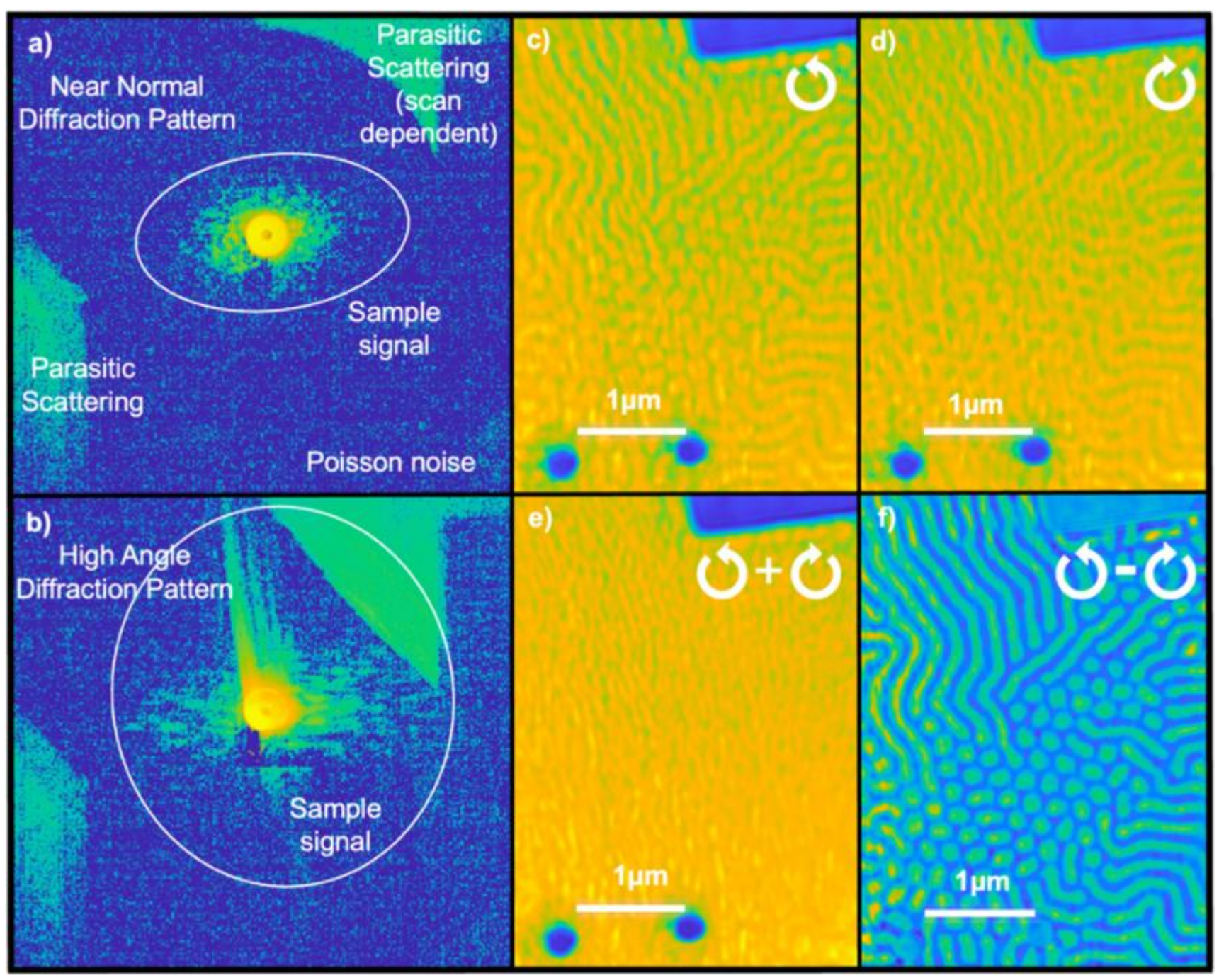

**Figure 5. Reconstructing the 2D images prior to scalar and vector tomography. a)-b)** Example diffraction patterns on a log scale. a) is a diffraction pattern from a ptychography scan taken at normal incidence, while b) was taken at -70º incidence. In addition to scattering from magnetic and electronic contrast, the pattern includes isotropic Poisson noise and structured parasitic scattering that varies with scan angle. **c)-d)** reconstructed optical density images taken with left and right circular polarization. **e)-f)** Images c) and d) are overlapped and aligned, then (e) summed to create the charge contrast projection, and (f) subtracted to create the magnetic contrast projection.

**2) XMCD Overlap:** For each angle in the tilt series, the two reconstructed OD images taken with left (Fig. 5c) and right (Fig. 5d) circularly polarized light were overlapped and registered to achieve maximum Mattes mutual information[53]. This metric enables registration between images with both equivalent contrast (structural signal) and opposite contrast (magnetic signal). Then, the sum of the two images was taken to extract the structural projection (Fig. 5e), and the difference to extract the magnetic projection (Fig. 5f). The quality of this overlap was confirmed by the fiducial holes having high contrast in the structural projections and zero magnetic contrast in the magnetic projections.

**3) Tomography:** Both the structural and the magnetic projections are taken as input to this tomography step. An accurate reconstructed 3D image requires both accurate registrations among all projections and algorithms that account for noise and limited sampling. We achieve these goals with a combination of pre-registration, iterative reconstruction, and registration correction.

3.1) Pre-registration: The independent ptychography reconstructions are not intrinsically registered to one another in this experiment, yet a strong initial guess for the registration of each projection image is necessary to achieve an accurate reconstruction with an extended object and limited rotation sampling. To register projections of the extended sample, we invert the structural projection signals by subtracting a uniform background, thus turning the fiducial holes into isolated pillar features. We track the center of mass (COM) of the four fiducial holes within our FOV in each projection image and treat these as the projected positions of four points rotating rigidly in 3D. We solve the inverse of this geometric projection to accurately estimate the in-plane translations (±5 pixels) and 3D rotations (± 5°) of each projection image.

3.2) Iterative reconstruction: To account for limited sampling in the experiment, our reconstruction algorithm took a real space iterative approach. We derived our algorithm based on RESIRE and RESIRE-V, for both scalar and vector tomography[54, 55]. Compared to closed-form solutions based on Fourier Slice Theorem[50,56,57], a real space iterative gradient approach generates much less missing-wedge effect under limited sampling.

3.3) Registration refinement: Due to the thin sample and rich magnetic contrast, the reconstruction quality is highly sensitive to registration accuracy. To further refine the registrations acquired in step 3.1), we iteratively determined the best shift and rotation registration after each tomographic reconstruction by minimizing the residuals between each measured projection image and the projection generated from the reconstructed object at the corresponding shift and rotation.

The refinement converged quickly in a few iterations, achieving sub-pixel shift and rotation accuracy.

More details of the full tomography reconstruction procedure are discussed in Supplementary Information.

**Acknowledgements**

This work was primarily supported by STROBE: a National Science Foundation Science and Technology Center under award DMR1548924. M.M. and H.K. acknowledge partial support by a Gordon and Betty Moore Foundation Grant No. 10784 for the algorithms that are robust to noise, the US Department of Energy, Office of Science, Basic Energy Sciences X-Ray Scattering Program Award DE-SC0002002 for the magnetic analysis, and an U.S. Air Force Office Multidisciplinary University Research Initiative (MURI) program under award no. FA9550-23-1-0281 for the low-contrast imaging technique. This research used resources of the Advanced Light Source, a U.S. DOE Office of Science User Facility under contract no. DE-AC02-05CH11231. S.A.M. acknowledges the support from the U.S. Department of Defense. The FIB preparation was carried out at the Facility for Electron Microscopy of Materials at the University of Colorado Boulder (CU FEMM, RRID: SCR_019306). J.Z. acknowledges the Quantum Center for the support received under No. INQC2600. We thank Eric Fullerton for valuable discussion regarding Fe/Gd skyrmions.

# Supplementary Information

# 3D Imaging of Complex Skyrmion Topologies in an Extended Sample

## 1. Videos of Tomographic Reconstructions

Videos may not play in document; links lead to a dropbox file.

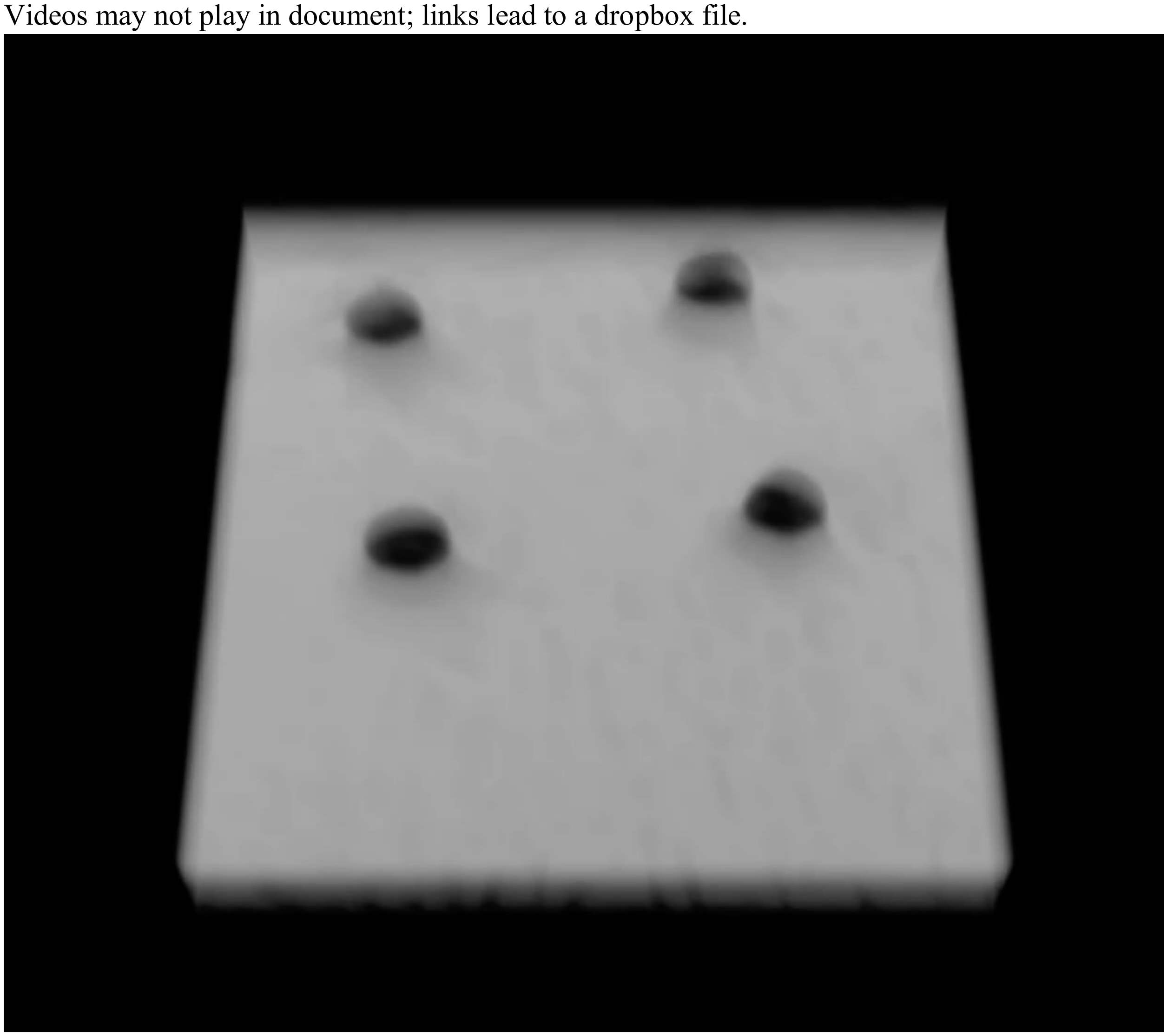

**1.1** ***Scalar Reconstruction Video*** Scalar reconstruction of the sample. The reconstruction rotates 360 degrees, showing the confinement of the electronic signal to an extended rectangle. The scalar reconstruction reflects sharp-edged fiducial holes and ripples in the SiN substrate. The conical shape of the fiducial holes through the depth of the sample is a result of the FIB procedure.

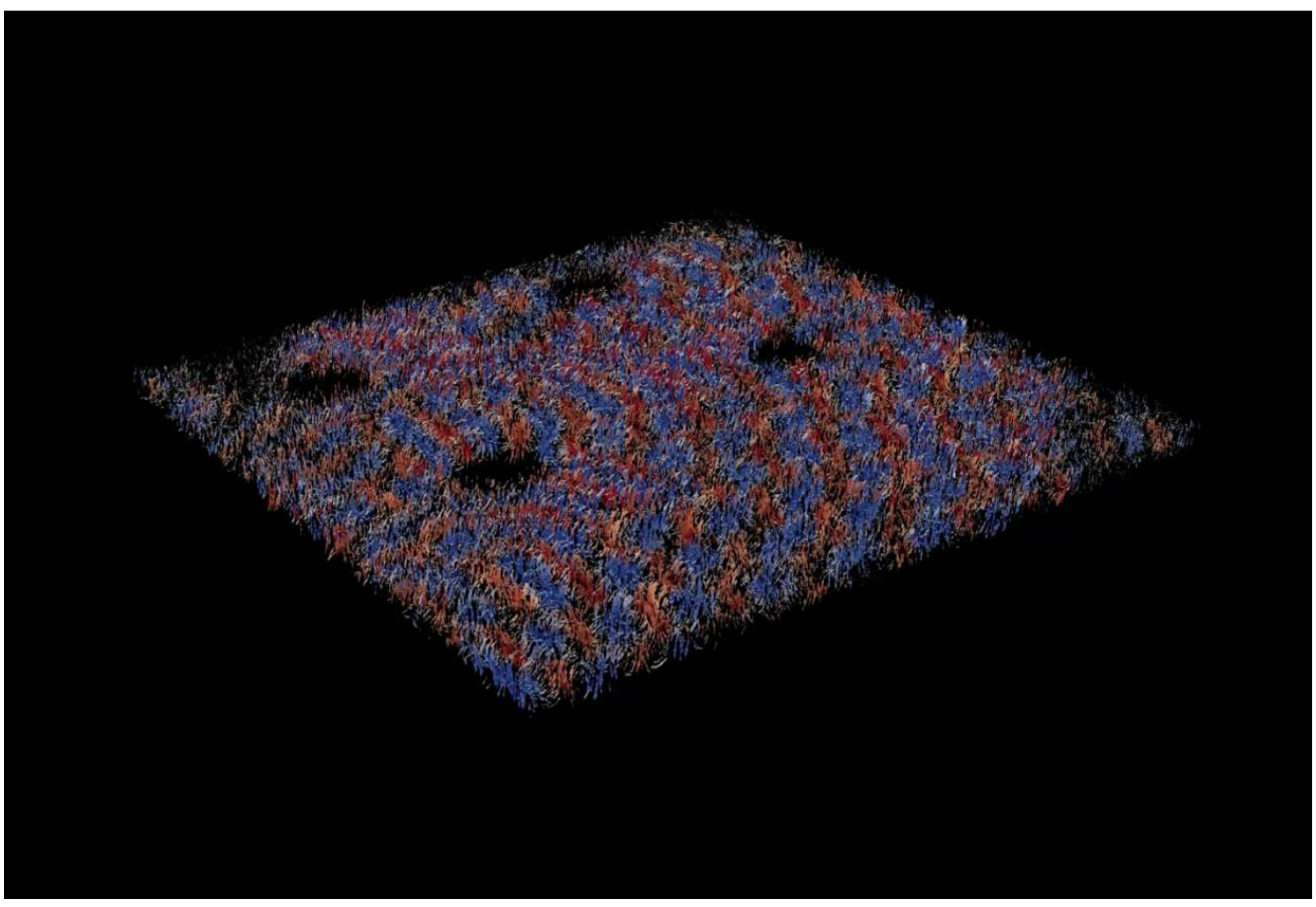

**1.2** ***Vector Reconstruction Video*** Vector reconstruction of the sample. The video begins with the full reconstructed FOV, then zooms in on a single skyrmion in the lattice. Arrows represent the reconstructed local magnetization, and the red-blue colormap indicated the magnetization z-component. The vector reconstruction is shown over an 88 nm thickness approximately corresponding to the actual depth of the magnetic multilayer. Reconstruction is downsampled for visual clarity.

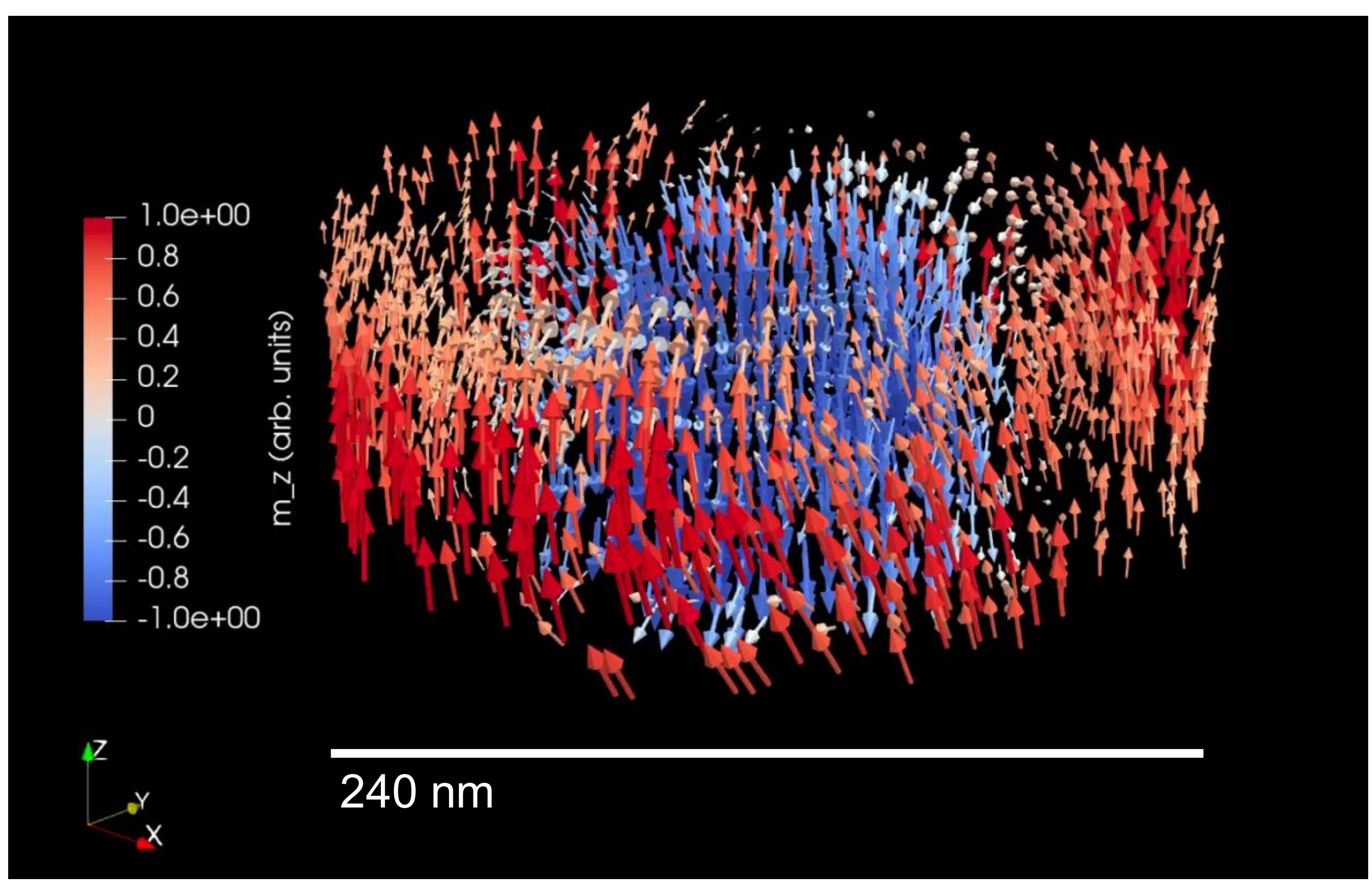

**1.3** ***Skyrmion Rotation Video*** Close up view of the rotating reconstructed skyrmion feature S20 (see Figure 2). Arrows represent the reconstructed local magnetization, and the red-blue colormap indicates the magnetization z-component. Reconstruction is downsampled for visual clarity.

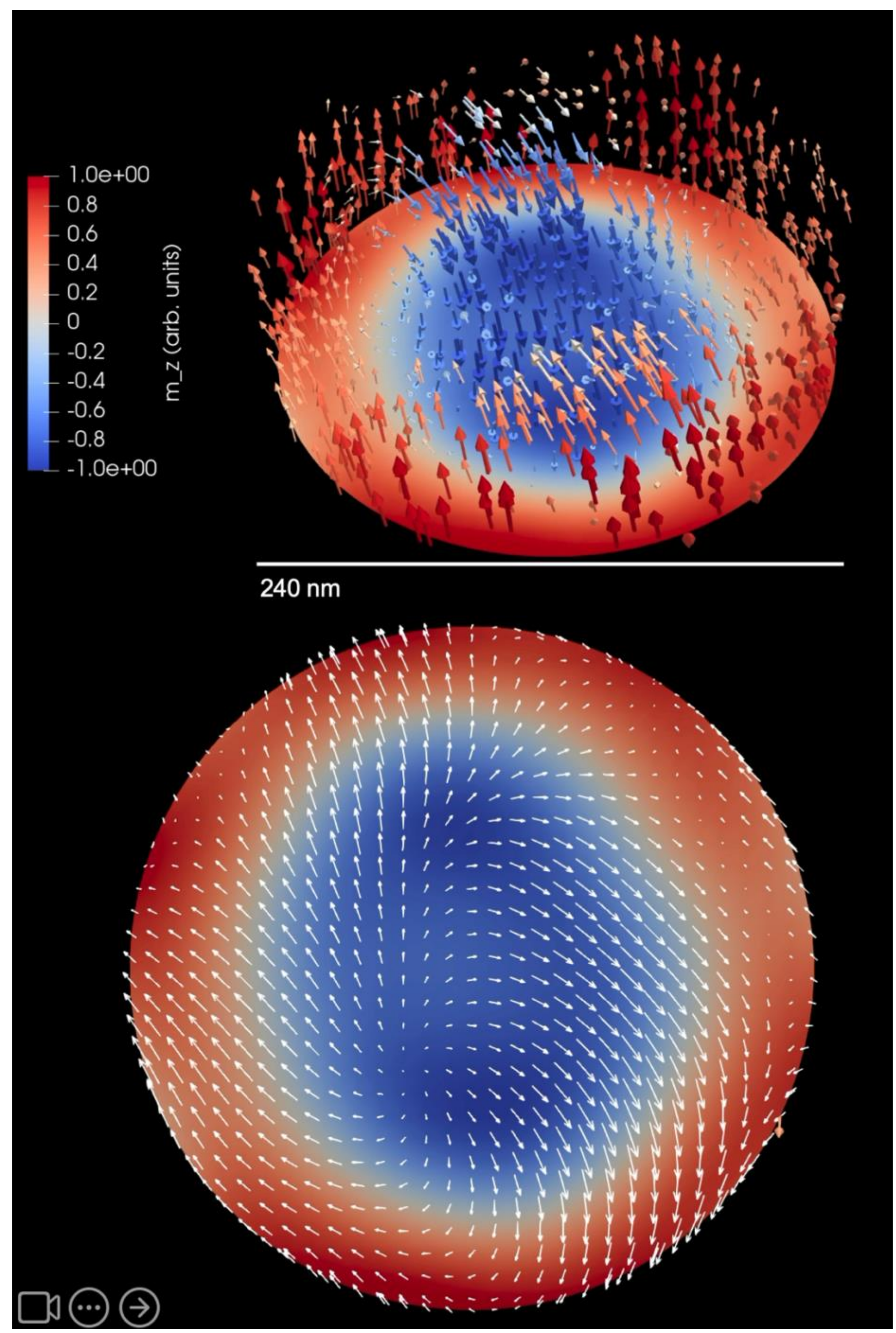

**1.4** ***Skyrmion Slice Video*** A rising cross-sectional slice through the reconstructed skyrmion feature S20 (see Figure 2). The upper view shows the position of the cross-section as it rises from the lower surface to the upper through the skyrmion. In the lower view, the white arrows show the local in-plane magnetization for every reconstructed voxel, while the red-blue colormap shows the magnetization z-component. The cross-sectional view demonstrates how the 2D slices transition from Néel type to Bloch and back to Néel.

## 2. Sample Synthesis and Preparation

The sample consists of an Iron (Fe)/Gadolinium (Gd) multilayer stack deposited via DC magnetron sputtering on a 100 nm silicon nitride substrate (Norcada). The Fe and Gd specimen was deposited by alternating layers of thickness 3.4 Å and 4 Å respectively, for a total of 120 bilayers. The resulting total thickness is approximately 92 nm. A 5 nm layer of Tantalum (Ta) was also added as a seed and cap to prevent oxidation. The SiN substrate itself developed a rippled texture sometime after deposition. This texture can be seen with SEM imaging (see SI Fig. 1) and is reconstructed in the ptychography experiment.

The sample was prepared and mounted for imaging using a 30 kV Ga focused ion beam (FIB) in an FEI Nova DualBeam. First, a 14.8 x 16.8 $\mu m^2$ square was cut from the membrane. This square was then attached to the central post of a 3 mm transmission electron microscopy (TEM) grid (Omniprobe, three-post copper lift-out grid, Ted Pella 460-2031-S) (SI Fig. 1b). Finally, fiducial marks were milled into the sample. Multiple sets of fiducial marks with different characteristics were included to provide multiple options for a suitable ROI in the ptycho-tomography experiment (SI Fig. 1c). The different features are as follows: 1) grids of holes ~300 nm in diameter, spaced by ~1 µm. The holes provide multiple isolated features in a single ROI without disrupting the skyrmion pattern of the sample. 2) An X-shaped hole, which could provide a more complex 3D structure to be used as a single fiducial mark in tomography. 3) Rectangular cutouts from the sample edge to provide the possibility of a partially isolated sample within a FOV of roughly 2x2 $\mu m^2$, and to provide clear transmission windows in a FOV that would also include other fiducial marks. To avoid ion beam damage during FIB milling, all the sample preparation steps, including cutting the large 14.8 x 16.8 $\mu m^2$ square and drilling the fiducial marks, were carried out from the SiN substrate backside of the sample.

After the fiducial marks were drilled and the sample attached to the grid, a magnetization procedure was performed to nucleate skyrmions. The sample was magnetized using a Montana Instruments Cryomagnet at room temperature. Magnetic fields were applied to the sample in the following sequence: 1) 400 mT at a 45-degree incidence angle to the plane of the sample 2) 0 field applied 3) 240 mT applied normal to the plane of the sample. The sample was positioned in the cryomagnet by manually attaching it to two different fixed mounts at 45 and 0 degrees. Identical magnetization procedures were performed on the thin film that the sample was lifted from, and the film was imaged using an MFM microscope to confirm the skyrmion phase.

The Omniprobe TEM grid was attached to a copper ring using silver paint to provide a full circular profile that could be clamped within the COSMIC beam line sample mount and manually rotated to an arbitrary angle. The copper ring is registered against the plane of the sample holder, but the 14.8 x 16.8 $\mu m^2$ square of the sample is not perfectly coplanar with that face, since only a single edge is attached to the Omniprobe grid.

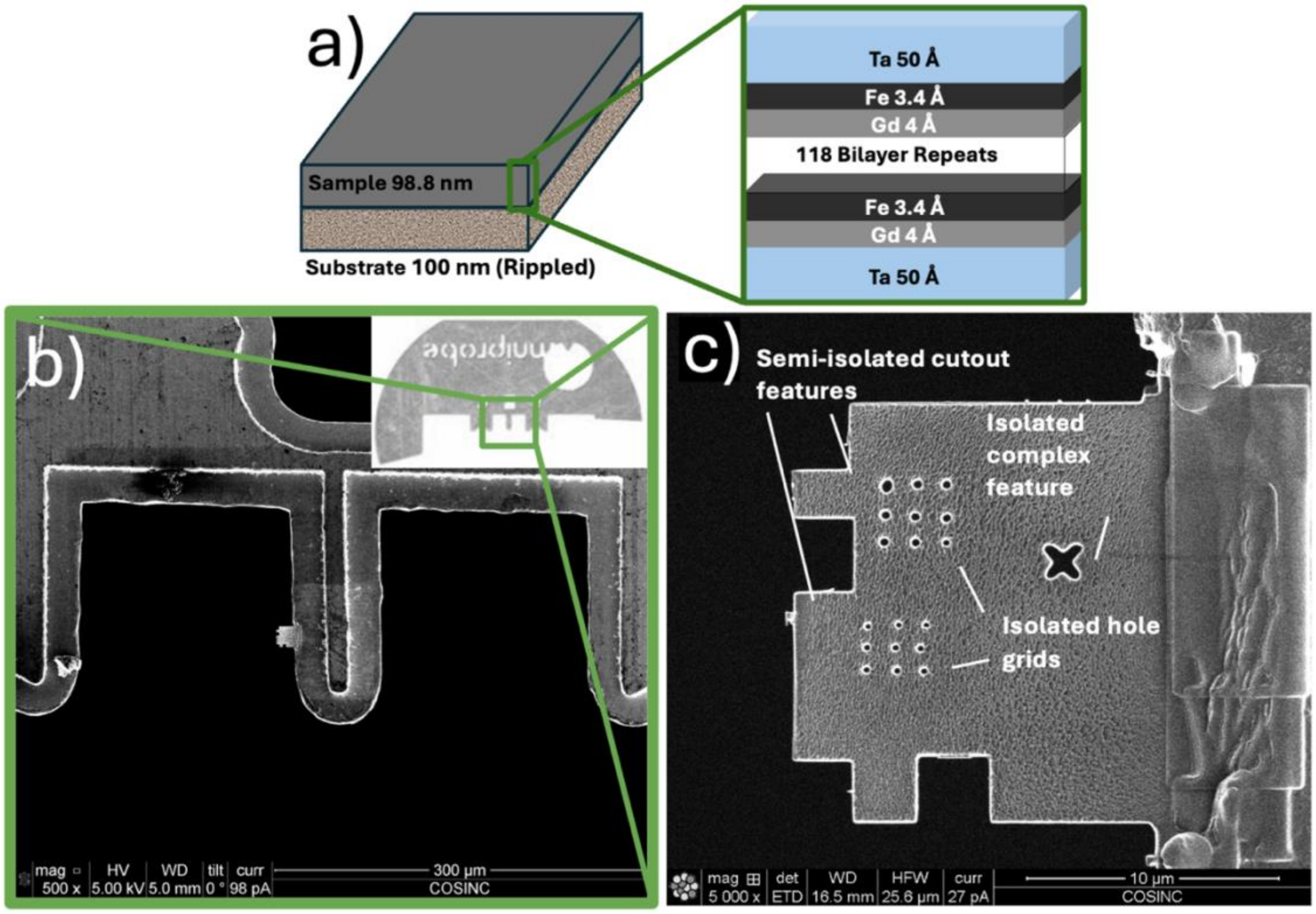

**SI Figure 1. a)** Composition diagram of the Fe/Gd multilayer. Fe and Gd are sputter-deposited in alternating layers of thickness 3.4Å and 4Å, respectively for a total of 120 pairs. A 50 Å Ta layer is used as a seed and cap. The structure is deposited on a 100 nm SiN thin film substrate. For more info on sample including micromagnetics simulations, refer to [35]. **b)** SEM image showing the 14.8x16.8 $\mu m^2$ square section of multilayer lifted out from a larger film by FIB. The section is mounted to a post on a copper omniprobe finder grid. **c)** SEM image of the sample liftout. This image shows various features drilled into the sample using the FIB as candidates for fiducial marks. The features included are holes in isolated grid patterns, a more complex isolated 'x' shape, and edge cutouts to provide partially isolated regions and clear transmission windows. This image shows the rippled texture of the SiN substrate on its backside.

## 3. Soft X-ray Ptycho Tomography Data Collection

The experiment was performed at the Coherent Scattering and Microscopy beam line (COSMIC) at the Lawrence Berkeley National Lab Advanced Light Source in Berkeley, California.

The sample is mounted on a TEM-style arm situated perpendicular to the beam path. The sample stage has approximately 70 µm of travel in the $x$ and $y$ directions perpendicular to the beam path, as well as 360 degrees rotation around the axis of the sample mounting arm. The experiment is performed at ambient temperature under high vacuum.

The detector is a CCD with area and 1024x1024 pixels at a distance of 120mm from the sample. The x-rays are focused using a 45 nm zone plate. STXM images are performed in focus, but ptychography images were taken with a defocus of 60 µm.

Left and right circular polarizations of the beam are used to extract both the scalar electronic signal and the vector magnetic signal according to the X-ray Magnetic Circular Dichroism (XMCD) effect.

This section describes the procedure for collecting the ptycho-tomography data series. Three tomographic tilt series were performed, with the sample manually removed and rotated in-plane in between each tilt series. The procedure for each tilt series was as follows.

**3.1** ***Initial measurements and calibration***

First, scanning transmission electron microscopy (STXM) images were taken to locate the sample. The working energy of 707 eV was chosen by performing an energy scan in a region of the sample with magnetic contrast and identifying the energy where contrast was maximized. Overview STXM images of the sample were taken to confirm the magnetic phase and identify potential ROIs.

Then, a calibration procedure was performed to ensure that the ROI on the sample was positioned close to the axis of rotation, to prevent the ROI from moving too far out of view with each rotation step. For this calibration procedure, the stage was rotated to positive and negative angles in increasing step sizes. The coarse $z$ positioning of the zone plate was adjusted such that the apparent 2D motion of the sample with rotation was minimized, and such that the motion was roughly symmetric around 0 degrees.

Then, the sample was rotated to extreme values in both the positive and negative direction to determine the angle limits for each tomography series. The stage has full range of motion, but due to the sample mounting geometry, the ROI would become occluded at different angles for each tilt series. The motion of the sample along the axis of the beam due to rotation was also recorded during this step by refocusing the zone plate onto a consistent edge of the sample at multiple rotations.

Then, with the rotation stage set to 0 degrees, the approximate in-plane orientation of the sample was measured by determining the angular orientation of a long straight-line feature on the copper TEM grid.

Then, the orientation of the sample normal vector with respect to the beam direction (i.e. the offset from a nominal 0 degrees) was estimated by refocusing the zone plate on sample features several microns apart in x and y and recording the z-positions of the zone plate.

These calibration and measurement steps were performed in between each tilt series after manually remounting the sample in the sample holder. In addition, the beam energy was reset to 707 eV at the start of each tilt series and once in the middle of the final tilt series to account for slight beam energy drift (~1 eV) over the course of the experiment.

**3.2** ***Tilt Series procedure***

Three tilt series were performed with angular ranges of -70 to 50 degrees, -60 to 60 degrees, and -60 to 60 degrees. The limits of these tilt series were chosen as the maximum angle before the ROI was occluded by the mounting setup. The series were each taken traveling from negative to positive angles in intervals of 2 degrees. Each series is taken in the order of increasing angle to avoid stage hysteresis. An interval of 4 degrees was used for the central 50 degrees of the scan for the latter two tilt series to reduce data collection time. The dataset is sparser when the sample is closer to flat and there is less change between subsequent projections.

The ROI was identified using STXM imaging at each angular step. The zone plate was refocused when necessary, more frequently at more extreme angles, and then set to a defocus of 60 μm for ptychography. The approximate spot size incident on the sample is 0.67 nm.

Two ptychography scans were taken at each angle in the tilt series, one each with left and right circularly polarized light. The polarization was switched between scans, with the scan parameters otherwise remaining the same. Each ptychography scan consists of a raster grid with step size 0.2 μm or 0.3 μm over an area of approximately 3x3 μm, with the dimensions changing with tilt angle and in-plane rotation to maintain the entire ROI within the rectangular scan grid. A random factor with a normal distribution, averaging to half the step size, was introduced to the scan positions to reduce gridding reconstruction artifacts that result from a perfectly periodic grid. A diffraction pattern was collected at each scan position by the CCD with an exposure time of 1 ms.

These ptychography scans were reconstructed for preliminary inspection on-site using an algorithm similar to SHARP[58]. Both polarizations were reconstructed, and a rough alignment was performed to inspect the XMCD images in real-time. This decoupled the rippled substrate texture from the magnetic domain texture for the purposes of quality inspection. If there were clear issues with focusing or shutter operation, the scan was repeated.

## 4. Soft X-ray Ptycho Tomography Reconstruction

### 4.1 *Data Preprocessing*

A small number of diffraction patterns were affected by a mechanical issue with the operation of the x-ray shutter. The real-time quality inspection did not identify these issues if only a small fraction of frames in a scan were impacted. These frames were later identified by their much higher number of counts and removed from the dataset.

Diffraction patterns were cropped from the detector size of 1024x1024 pixels to 512x512 pixels. The highest frequency scattering from the sample was still contained within the smaller field of view, and this cropping reduced the impact of high frequency parasitic noise while also speeding up reconstruction time.

### 4.2 *Ptychographic Reconstruction*

Ptychographic reconstruction was done using the Partial Coherence and Monochromatization Algorithm with Noise, or PaCMAN[45]. The reconstructions were done with one object mode and three probe modes. No monochromatization or multi-wavelength correction was done. The probe was initialized as the probe reconstructed at the COSMIC beamline, while the object was initialized as a matrix of ones. We terminated the algorithm after 200 iterations, with probe updates beginning at iteration 50 and parasitic scattering updates beginning at iteration 20. We used a size parameter of $r = 0.6$.

### 4.3 *OD Calculation*

Each ptychographic reconstruction was converted from a transmitted intensity ($I$) image into an optical density image. A background value was calculated for each image by taking the histogram of a region containing both sample and vacuum, but no fiducial holes due to their varying transmission across the tilt series. This histogram has two peaks, representing sample and vacuum. The background value for incident intensity ($I_0$) was chosen as the peak location of the vacuum region histogram. The OD image was calculated as $OD = log(I_0/I)$.

**4.4 *XMCD Alignment***

Registration of the left and right circularly polarized images to each other was performed using the Matlab *imregtform* function on a cropped region of the reconstructions centered on the fiducial holes. The resulting sum and difference images were inspected for quality by ensuring that there were no high contrast areas around the holes in the magnetization image (see Fig. 5).

In this experiment, the polarizations are labelled 0 and 1, and it was not known which state corresponded to left vs. right circularly polarized light. The 1 state was consistently subtracted from the 0 state throughout the dataset, but there is therefore an ambiguity in the XMCD calculation which amounts to a sign flip in the component of the magnetization parallel to the beam. This results in an overall flip of the magnetic distribution ($\boldsymbol{m} \rightarrow -\boldsymbol{m}$). Notice that this sign flip will switch the sign of *N* in the 2D topology calculation, as can be seen in Eq. 1,2. However, this will not change the topological nature of skyrmions (i.e. to antiskyrmions), as a skyrmion is defined by its topological charge together with the direction of the skyrmion core, which also switches when $\boldsymbol{m}$ is flipped. Therefore, the features are still identified as skyrmions rather than antiskyrmions despite this ambiguity.

**4.5 *Absorption Correction***

The energy of the x-ray beam was tuned to maximize contrast at the beginning of the beam time and was reset to the resonant energy at the beginning of each tilt series plus one additional time during the third tilt series. The slight energy drift between these tunings (< 1eV) results in a reduction of the XMCD contrast as the tilt series proceeds. SI Figure 2 shows that the root mean squared (RMS) value in the XMCD signal drops as the beam energy drifts, rather than following the expected curve due to sample tilt.

To correct for this drop in contrast, we assume that: 1). the strength of the XMCD absorption signal from a thin film is approximately proportional to the effective thickness along the beam direction for moderate tilt angles, and 2). the absorption will drop linearly for small shifts in beam energy away from resonance. To implement this correction, we calculate the root mean square (RMS) values of each projection, $\mathrm{RMS}_{\mathrm{OD},i}$ (SI Fig. 2b), and model the signal as

$$\mathrm{RMS}_{\mathrm{OD},i} = \frac{1}{2} A \cdot t_{\mathrm{eff}} [1 - p \cdot (E_i - E_0)], \qquad \text{[SI.1]}$$

where $A$ is the average magnetic OD strength, $t_{\mathrm{eff}}$ is the transmission distance through the sample for the *i*-th tilt angle, and $p$ is the linear decay rate associated with the recorded energy drifts ($E_i - E_0$) (SI Fig. 2a). To avoid inaccuracy of such approximation at high tilts, only projections with polar angle of plane normal < 30° are used for fitting. Finally, each XMCD image is corrected by dividing by the factor $[1 - p \cdot (E_i - E_0)]$ (with the fitted decay rate $p$ and recorded energies plugged in) prior to vector tomography reconstruction.

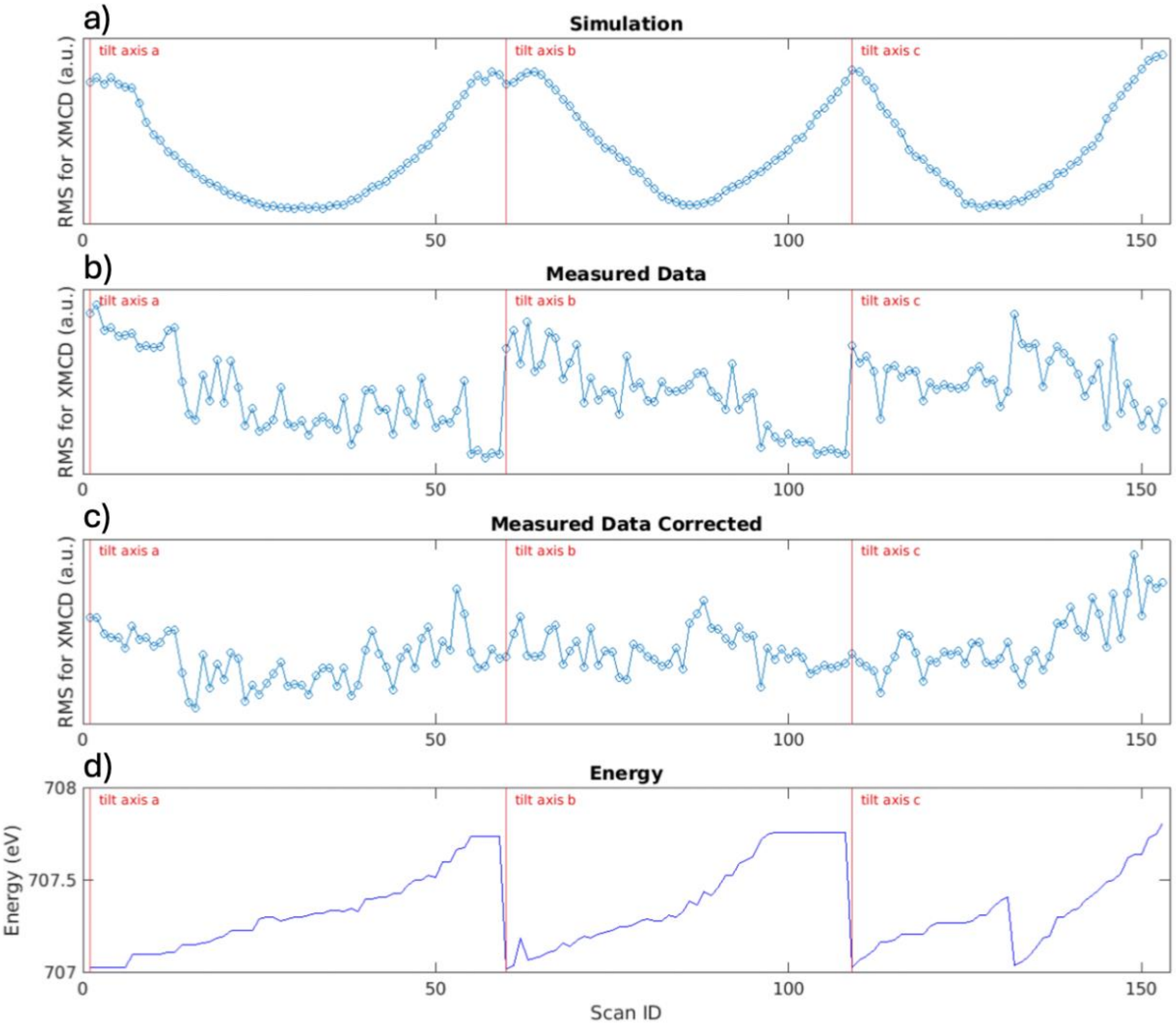

**SI Figure 2.** A comparison between **a)** the expected RMS of the XMCD signal for each projection, **b)** the measured RMS, and **c)** the rescaled measured RMS data. **d)** shows the recorded beam energy over the course of the experiment. As the beam energy drifts away from the resonant peak (707 eV) the RMS of the data drops, which is then corrected by rescaling.

### 4.6 *Tomographic Reconstruction*

An accurate tomographic reconstruction from structural and magnetic projection images involves a coarse pre-registration, an iterative reconstruction algorithm, and a registration refinement mechanism.

**4.6.1) Pre-registration**: As described in Methods.

**4.6.2) Iterative structural reconstruction:** The data fidelity loss function used in RESIRE is defined as the sum of squares of the residual between measurement and reconstruction:

$$\mathrm{L}(m) = \frac{1}{2}\sum_{\theta}\|A_{\theta}m - b\|_2^2 \quad \text{[SI.2]}$$

Here $m$ denotes the reconstructed object field (either a scalar or vector field), $A_{\theta}$ is the linear forward rotation and projection operator of the object that generates a projection image at a 3D rotation labeled by $\theta$, each of which labeled by three Euler angles in the intrinsic "*yxz*" order. Then, an implementation of the forward operator $A_\theta$ and its transpose $A_\theta^T$ enables various gradient approaches to minimize the loss function.

**4.6.3) Registration refinement:** Since vector tomography with solely longitudinal signal is ill-posed, only the structural signal from the scalar tomography is reliable to be used for registration refinement, and the registrations for magnetic projections are adjusted accordingly.

**4.6.4) Iterative magnetic reconstruction:** Since the magnetic and structural projections are both created from the registration between left and right circularly polarized OD images, they are inherently registered to one another. Using the same iterative logic, as well as the refined registration obtained from the previous steps, we then perform the vector version of iterative tomography, RESIRE-V, with the magnetic projections[55]. Fig. 3 in the main text shows the results of this final step. The reconstruction is performed with a solution space slightly thicker than the sample (and substrate for structural reconstruction) to account for the missing wedge effect caused by incomplete sampling of the tilt series[25].

**4.7 *Topological Analysis***

To perform the quantitative topological analyses mentioned in the main text, we numerically calculate the skyrmion charge density and emergent magnetic field from the reconstruction. To ensure numeric stability, we use the solid angle method[59-61].

In addition, to calculate the Hopf index, we also need the vector potential for the emergent magnetic field. While nontrivial to compute in real space, the vector potential can be conveniently computed in Fourier space given the Coulomb gauge. The analytical formula for the Hopf index in Fourier space is:

$$H = \int_V \boldsymbol{F}^*(\boldsymbol{k}) \cdot \frac{-i\boldsymbol{k} \times \boldsymbol{F}(\boldsymbol{k})}{2\pi k^2}\, d^3\boldsymbol{k}. \qquad \text{[SI.3]}$$

For numerical computation, to ensure zero imaginary part of the calculated Hopf index, we implement the following formula:

$$H = \sum_{i,j,k} \frac{i\boldsymbol{k}_{i,j,k} \cdot \left(Im\left[\mathcal{FFT}\{\boldsymbol{F}\}_{i,j,k}\right] \times Re\left[\mathcal{FFT}\{\boldsymbol{F}\}_{i,j,k}\right]\right)}{\pi N_x N_y N_z\, k_{i,j,k}^2} \qquad \text{[SI.4]}$$

Here $\mathcal{FFT}\{\cdot\}$ denotes the fast Fourier Transform, $\boldsymbol{k}$ is the wave vector conjugate to spatial coordinates, $N_x, N_y, N_z$ are the number of points along the *x,y,z* dimensions, and *i,j,k* are the indices along the *x,y,z* directions in *k*-space grid. To ensure enough sampling precision during the numerical Riemann sum, the emergent magnetic field is zero-padded to 10x its original size in all dimensions before taking the fast Fourier Transform.

**4.8. *Skyrmion Shape Calculations***

Calculations were performed as described in the main text to quantify the skyrmion size and shape. The skyrmion radial size as defined from the $m_z = 0$ contour in equation [3] for all 24 skyrmions is shown in SI Figure 3, along with the depth-dependence of the skyrmion size. The skyrmion domain wall radial parameter is largest in the bulk with a median size of 83nm, and is smaller by an average of 2nm at the surfaces.

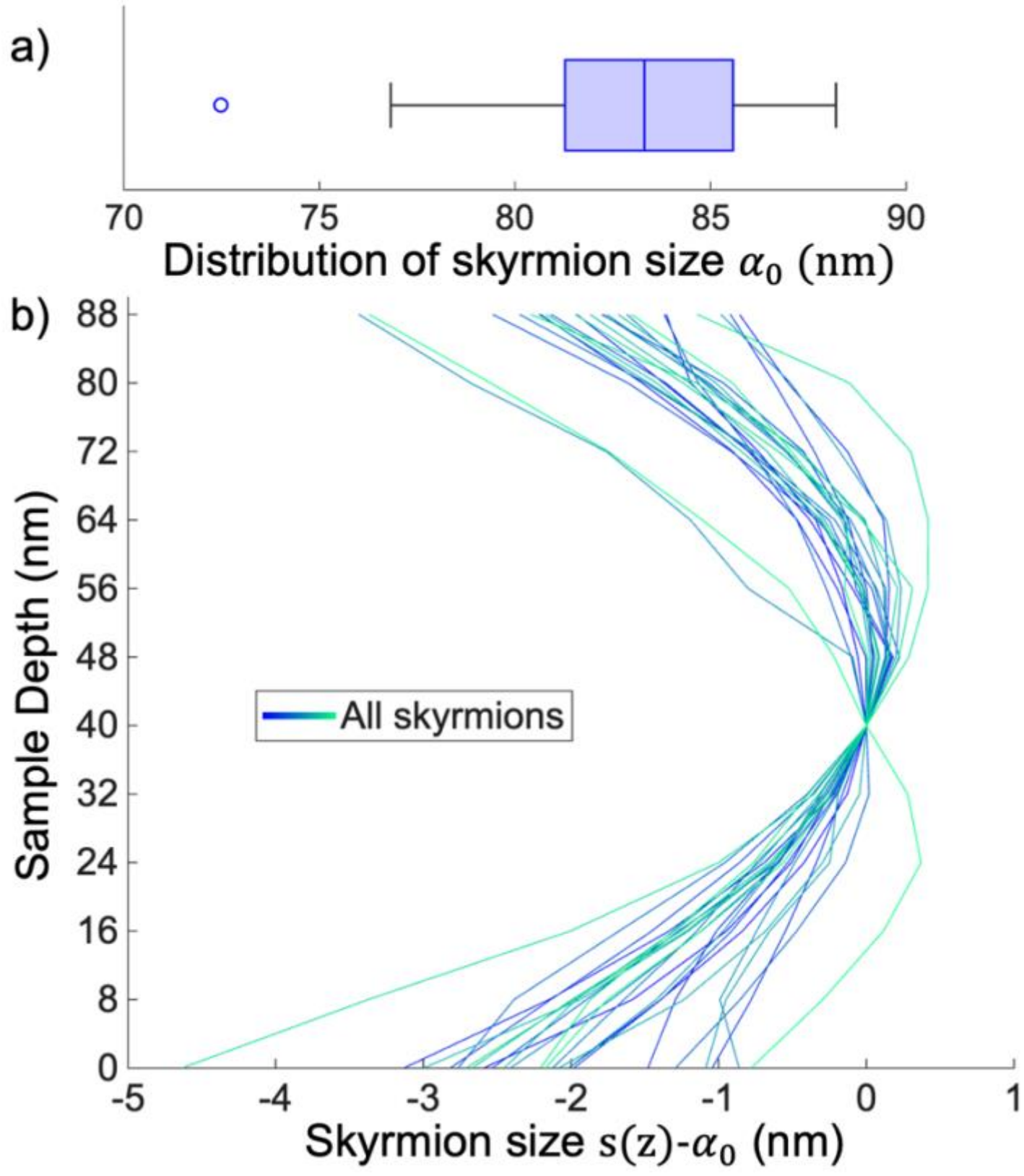

**SI Figure 3** Size characteristics of all skyrmions. **a)** shows a box chart of the characteristic skyrmion size parameter $\alpha_0$ for all 24 skyrmions. In **b)**, $\alpha_0$ is subtracted from the depth dependent skyrmion size $s(z)$ to show the change in skyrmion size as a function of depth.

### 4.9 *Effect of Sampling Bias*

Due to the limits of mounting and tilting an extended sample, the maximum tilt angle is limited to 60°-70°, rather than a full 90° degrees. This incomplete span of angular sampling (see solid angle sampling inset in Fig. 1) leads to a sampling bias due to the missing wedge effect[50]. In a typical scalar tomography setup, only a single tilt series is performed, and the problem is reduced to slices of 2D tomographic reconstruction problems[62-64].

In addition to the missing wedge effect, since the magnetic projection images show the vector magnetization field projected along the beam direction, this effect also leads to biased sampling of the strength of the three components $(m_x, m_y, m_z)$ of the magnetization field. Specifically, this sampling bias causes a suppression of the in-plane component of the vector reconstruction compared to the out-of-plane component, and a slight suppression of the 30° direction compared to the 120° direction. This suppression has a limited impact on the local topological calculations for the reconstruction, as described below.

The helicity angles are locally distorted, but this distortion is averaged out across the domain walls when we report the helicity for a skyrmion feature.

When calculating the topological charge, the suppressed in-plane magnetization components will distort the vector directions when $\boldsymbol{m}$ is normalized to calculate $\rho$, and will tend to artificially concentrate $\rho$ more tightly around the domain walls. Thus, it is preferable to define the skyrmion domain wall in terms of the profile of $m_z$ independent from the other components rather than the alternative definition based on the distribution of topological charge. Note that, though the local distribution of $\rho$ is distorted, the total topological charge integration is unaffected because the distortion still belongs to a smooth deformation protected by topology and the magnetization vectors still complete a full twist across each skyrmion.
In summary, this bias effect impacts the reconstructed direction and magnitude of magnetization vectors, but does not change features such as skyrmion distribution, size, integrated topological charge, or helicity.

**5. Micromagnetic Simulations**

In addition to the simulations reported in [35], micromagnetic simulations were performed to predict the magnetic texture in the presence of fiducial holes, and to verify the performance of the vector tomography algorithm for this sample. The simulation was performed using the GPU-based MuMax3[65,66] finite differences in time domain simulation package. For the simulations an amorphous Fe/Gd film was assumed with the exchange stiffness $A_{ex}$ = 5e-12 J/m, the saturation magnetization $M_{sat}$ = 4e5 A/m and a uniaxial anisotropy $K_{u1}$ = 4e4 J/m$^3$ with anisotropy vector perpendicular to the film surface, reflecting the parameters in [43]. The simulation uses a series of energy minimizations following the experimental magnetizing procedure to obtain the final result.

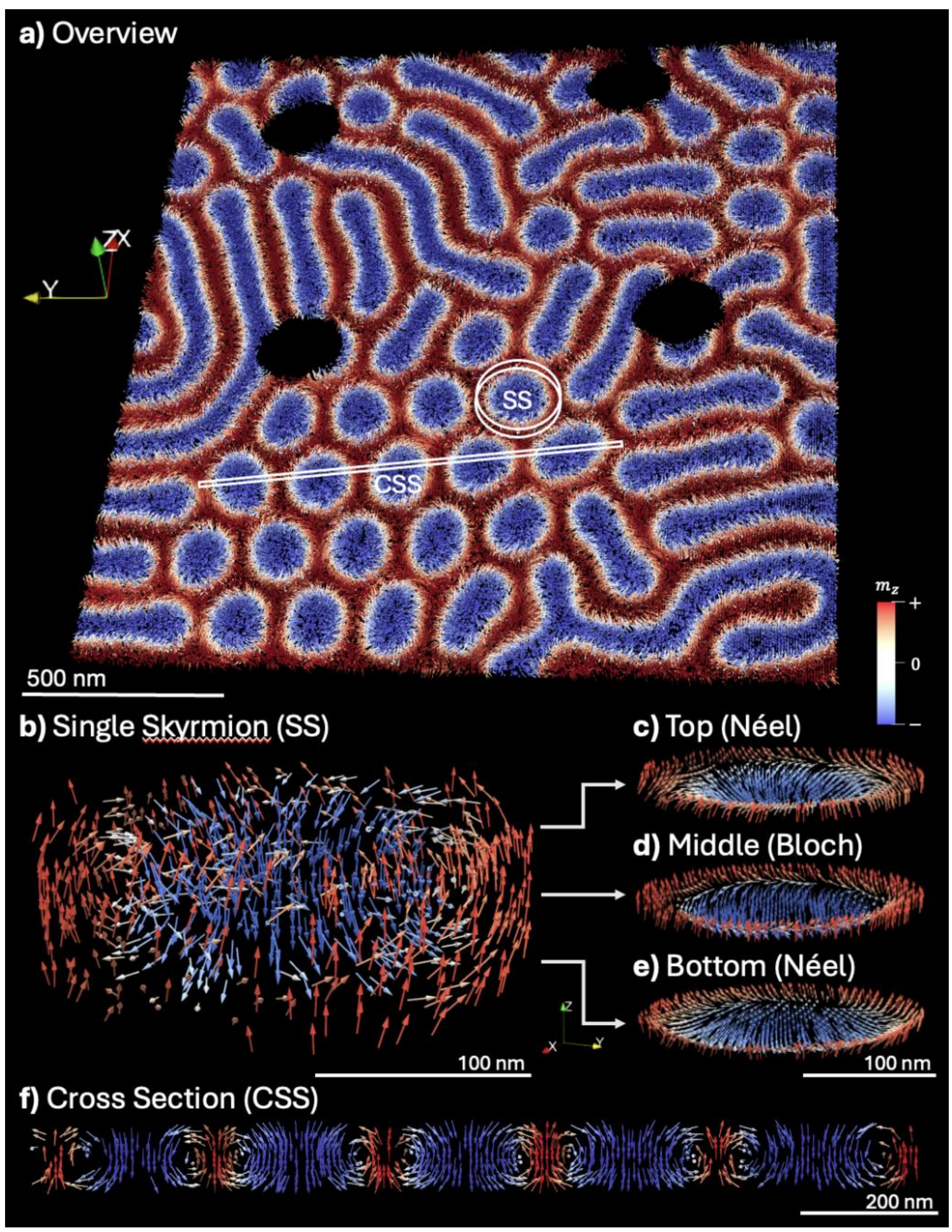

**SI Figure 4.** 3D view of the simulated magnetization distribution, with similar layout to Fig. 3 and Fig. 4 in the main text.

## 6. Verification of Reconstruction

To verify that our reconstruction is showing the real texture in our sample rather than the results of artifacts, we performed the reconstruction procedure on simulated datasets with various sources of noise present.

We generated a virtual sample from the micromagnetic simulation (see SI section 5) with an added rippled scalar substrate. We generated projection datasets from this sample via direct numerical projection. We did not simulate the ptychography process because its robustness and ability to reduce artifacts in this dataset were previously evaluated in [45].

The sources of noise added to the simulated datasets were as follows:

1) Limited sampling. We generated projections in the same sampling scheme as our original data to replicate the missing wedge effects caused by incomplete sampling.
2) Normalization error. Some error is introduced during the OD calculation. Beam fluctuations and challenges with ptychographic reconstruction of a vacuum region result in noise in the chosen transmission value. We introduce a similar error in the simulated projections.
3) Shifting beam energy. The reduced contrast over the course of the tilt series from shifting beam energy explained in SI section 3.5 is applied to the projections. (See SI Fig. 2)
4) Angular misalignment. Discussed in the main text. (See Fig. 5 and SI Fig. 5)

The reconstruction generated from the simulated data with added noise from sources 1-3 has minor deviations from the ground truth simulation, but no major alterations to the magnetic texture and topology.

Angular misalignment does impact the magnetic topology, as outlined in the main text. SI Fig. 5 shows that the reconstruction of a noise-free simulated sample contains no topological defects within the fully sampled region (covered by all projection FOVs and within scalar support), but that an added angle offset error results in topological defects within this region.

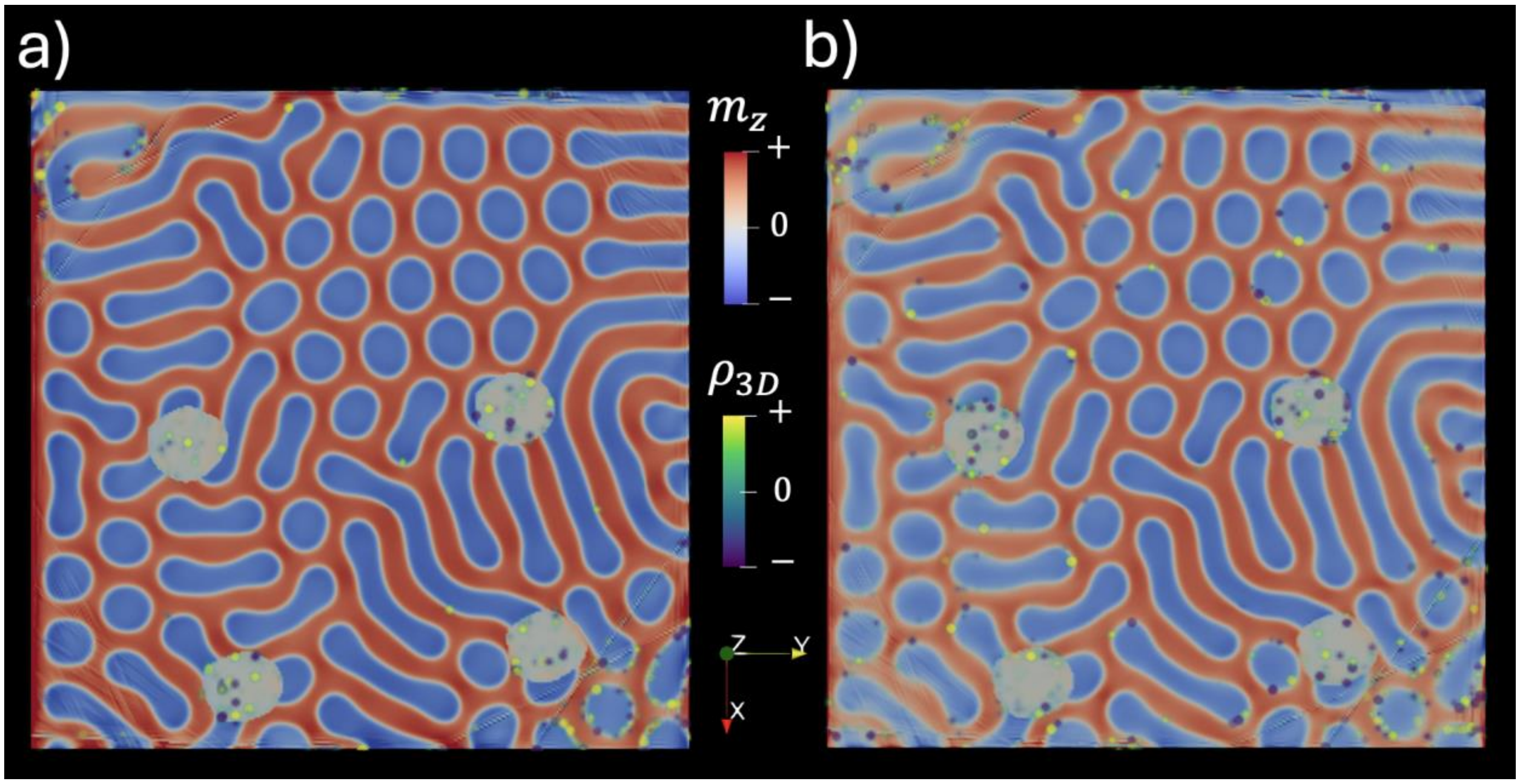

**SI Figure 5.** A top-down view of the vector reconstruction and topological charge of the simulated data **a)** with no error added and **b)** with angle offset error added. Blue-red color indicates $m_z$, the *z* component of the reconstructed magnetization field, of the central *xy* slice. Violet-yellow color indicates 3D topological charge density calculated by Eq. 3,4 in the main text. Topological defects within the fiducial holes and in the undersampled corners can be disregarded, but angular misalignments introduce topological defects into the central region of the reconstruction.

## 7. Fourier Shell Correlation

To quantitatively analyze the spatial resolution of the tomography, we carry out a Fourier Shell Correlation (FSC) calculation on both the structural and magnetization reconstructions following a similar logic to a previous work[30]. We divide the projection images after coarse registration into two datasets by even and odd indices and carry out separate registration refinement procedures and independent tomography reconstructions. Then, a FSC calculation is performed between the two sets of reconstructions. SI Fig. 6 shows the FSC for the structural reconstruction and the norm of *x,y,z* components of the magnetization reconstruction. In each plot, the spatial frequency is plotted up to the Nyquist frequency limit of 2 voxels. The FSC value of both structural and magnetic reconstructions remains above the commonly quoted threshold of 0.143 out to the Nyquist frequency, confirming that our method accurately reconstructs the sample.

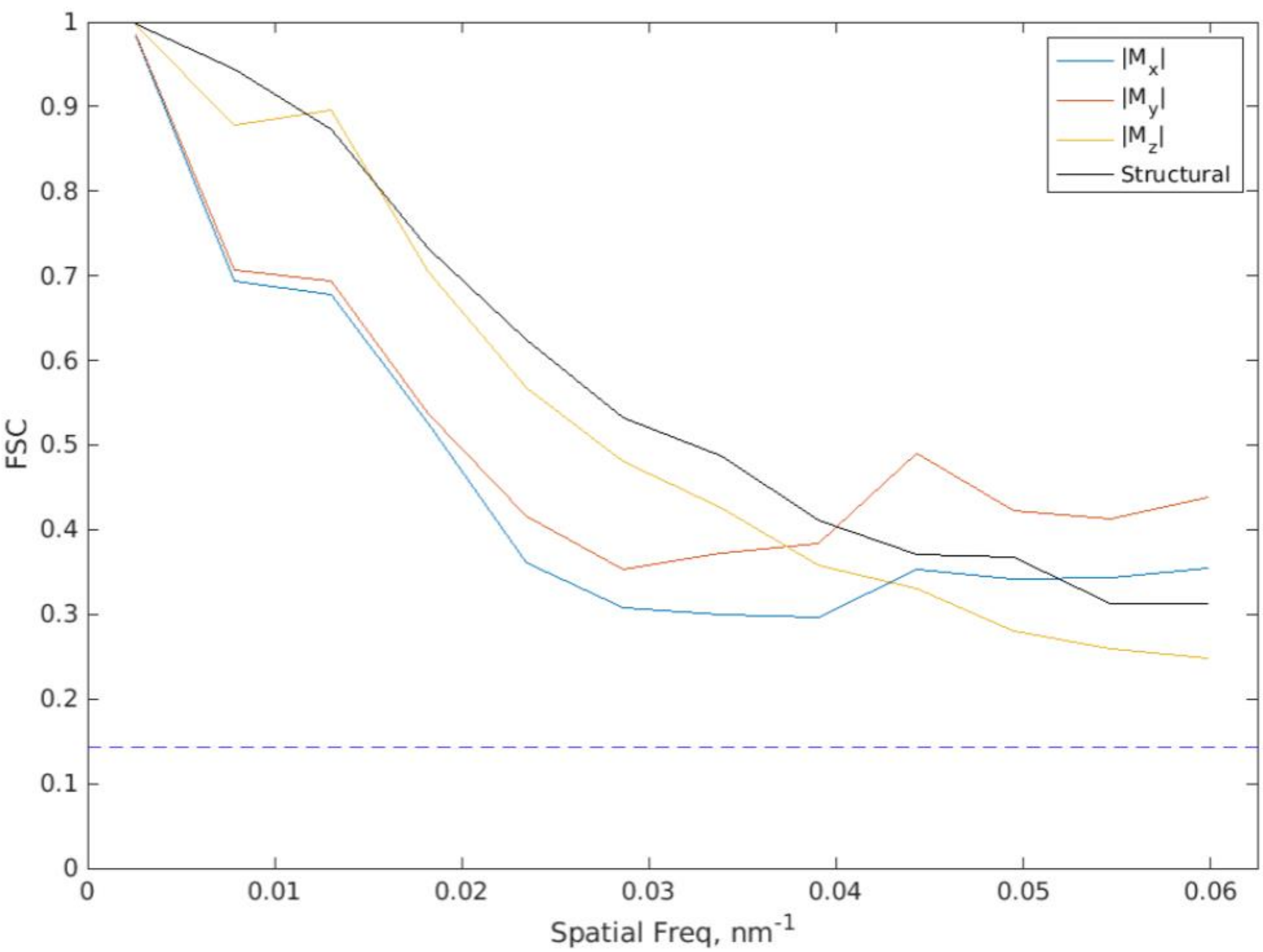

**SI Figure 6** Fourier Shell Correlation Analysis for the structural reconstruction and the individually normalized components of the magnetization reconstruction. The blue dotted horizontal line shows the 0.143 threshold.